\documentclass[11pt]{article}

\usepackage{geometry}
\author{Tomislav Grospić\\{\small\texttt{grospic@gmail.com}}}
\date{July 2026}

\usepackage[utf8]{inputenc}
\usepackage[T1]{fontenc}
\usepackage{lmodern}

\usepackage{amsmath}
\usepackage{hyperref}
\usepackage{enumitem}

\newtheorem{definition}{Definition}[section]

\title{Consensus as Collapse Policy:\\
Communication Evidence, Horizons, and Prefix Decisions}

\usepackage{tikz}
\usepackage{array}
\usepackage{pdflscape}
\usetikzlibrary{arrows.meta,positioning,shapes.misc}
\newcommand{\denote}[1]{\left\lbrack\!\left\lbrack #1 \right\rbrack\!\right\rbrack}
\newcommand{\specdesc}[1]{\vtop{\hbox{\begin{tabular}[t]{@{}l@{}}#1\end{tabular}}}}
\newenvironment{specdisplay}
{\[
\small
\renewcommand{\arraystretch}{1.05}
\begin{array}{r@{\quad}c@{\quad}l}}
{\end{array}
\]}

\begin{document}
\maketitle

\begin{abstract}
Consensus protocols are usually specified by their terminal artifact: a decided value, replicated log, or finalized prefix. This output-first view hides the communication-derived evidence that makes such artifacts safe: messages, votes, certificates, causal dependencies, equivocation evidence, timeouts, and local views. This paper makes that carrier explicit. We model distributed execution as an order-2 evidence state induced by communication and read classical consensus outputs as order-1 projections of that state.

Under this view, consensus protocols can be compared as collapse policies. A protocol specifies which evidence is legitimate, which finite horizon it inspects, when it projects a communication-induced evidence state into a value or prefix, and how it repairs or defers collapse when the visible evidence is insufficient. This separates legitimacy from collapse: quorum intersection, committee eligibility, checkpoint votes, sampling confidence, estimator safety, or DAG support can make evidence decision-grade, while the collapse policy determines when preserved distinctions are allowed to stop mattering.

The impossibility lineage supports the same distinction. FLP constrains deterministic guaranteed collapse to a terminal decision under full asynchrony with one crash failure; set agreement exposes the width of the output carrier; topological distributed computing characterizes when a history/view carrier admits a structure-preserving map to an output carrier. The contribution is therefore not a new impossibility theorem or a replacement for protocol-specific proofs, but a denotational specification framework: consensus is collapse under evidence.
\end{abstract}

\section{Introduction}

Consensus is commonly presented as the problem of making distributed processes agree on a single value. That definition is correct for the classical task: agreement, validity, and termination are stated over a terminal decision \cite{fischer-lynch-paterson-1985}. But concrete protocols do not begin with a decided value. They begin with proposals embedded in communication. Processes exchange messages, form local views, collect votes, detect equivocation, wait for rounds or waves, build certificates, follow leaders, and sometimes defer decisions through fallback rules.

This paper studies the structure that appears before the terminal decision. The central claim is simple:

\begin{quote}
Nontrivial consensus is lawful projection from communication history: protocols differ less in their terminal artifact than in their policy for making accumulated communication evidence legitimate, visible, and stable enough to collapse into an order-1 output.
\end{quote}

In many protocols, legitimacy is supplied by an intersection or threshold condition. In crash-fault protocols such as Paxos and Raft, majority quorums intersect, so a later leader can recover the value or log prefix that may already have been made durable \cite{lamport-paxos-1998,ongaro-raft-2014}. In Byzantine quorum protocols, the boundary is stronger: with $n=3f+1$, any two quorums of size $2f+1$ intersect in at least $f+1$ validators, hence in at least one correct validator. Weighted quorum systems, resource-weighted systems, and flexible quorum systems change the unit being counted or the pairs of quorums that must intersect. Federated quorum slices, private committee sortition, stable-prefix rules, finality gadgets, and repeated sampling change how decision-grade evidence is carried, but they still give the collapse rule a structured carrier to read.

The real variation lies in the carrier and its closure policy. Protocols choose how much communication-induced evidence to preserve before output, how they compress that evidence, when they stop waiting, and what they do when the visible evidence is insufficient. The same pattern is already visible in the original Byzantine Generals Problem: oral-message recursion and signed messages are two ways of preserving enough communication evidence before collapsing to a common order \cite{lamport-shostak-pease-1982}. Leaders, rounds, waves, quorum-certificate chains, committee credentials, checkpoint links, sampling windows, direct and indirect rules, timeouts, anchors, and random leaders are later mechanisms for controlling this collapse boundary.

\paragraph{Communication-history reading.}
In nontrivial distributed consensus, the primary carrier is not the terminal value. It is the communication-induced evidence state from which the value becomes justified. When the problem is framed output-first, the field studies increasingly elaborate ways to make the collapsed artifact safe. When framed communication-first, the same mechanisms become policies for preserving, compressing, delaying, or repairing the collapse from evidence state to prefix.

In this sense, nontrivial distributed consensus is necessarily communication-mediated. If different components may begin with different information, then agreement satisfying a nontrivial validity condition must depend on some interaction structure by which information becomes shared, constrained, or made observable. The paper uses communication in this broad concurrency-theoretic sense, not merely as explicit network transmission.

\paragraph{Lamport's starting point.}
This communication-first view is already present in Lamport's happened-before relation \cite{lamport-time-clocks-1978}. Lamport does not begin with a single global sequence. He begins with distributed events and the partial order induced by process-local order and message send/receive causality. Logical clocks then provide a way to extend that partial order into a total order when a system needs one. In the terminology used here, the causal communication history is the carrier; the total order is a projection. Consensus protocols repeat the same semantic move: they construct communication evidence and then project it to a value, log, block, or prefix. Protocol presentations often foreground the projected artifact, but the semantic work is carried by the communication evidence.

\paragraph{Order terminology.}
We use ``order-2'' for the communication carrier: the evidence state induced by labelled events together with causal and evidential relations among them. We use ``order-1'' for the terminal artifact exposed by the usual consensus abstraction: a value, a linear log, or a finalized prefix. The terminology is semantic rather than hierarchical: order-2 objects remember relations between changes; order-1 objects expose the chosen state or prefix after those relations have been compressed.

\paragraph{Closure criterion.}
Meaning closes at a carrier when the correctness question can be answered from that carrier without consulting distinctions preserved only at a higher order. Thus collapse from order-2 history to an order-1 prefix is valid only when the evidence visible at the chosen horizon makes the discarded distinctions irrelevant.

\paragraph{Thesis.}
Consensus protocols have accumulated many protocol-specific vocabularies. Lamport clocks speak in happened-before order and logical timestamps; Paxos in ballots, promises, and accepted values; Raft in terms, leaders, and log replication; PBFT in phases and views; Stellar in quorum slices and federated ballots; HotStuff in quorum-certificate chains; Tendermint and Streamlet in rounds, locks, notarizations, and finalized chains; Algorand in VRF sortition and committee certificates; Ouroboros in slots, stake, leader eligibility, and stable prefixes; GRANDPA and Casper FFG in finality votes and justified checkpoints; Avalanche/Snowman in repeated sampling and confidence thresholds; Jolteon and Ditto in fast paths and asynchronous fallback; CBC Casper in estimates, justifications, protocol states, and safety oracles; DAG-Rider in DAG waves and local interpretation; Hashgraph in gossip-about-gossip, witnesses, fame, and virtual voting; Aleph in units, posets, and common-coin ordering; Cordial Miners in blocklace, ratification, and final leaders; Mysticeti in direct and indirect slot rules. These names matter operationally, but semantically they often play the same role: they determine when a communication-induced evidence state is sufficiently legitimate, sufficiently visible, or sufficiently old to be collapsed into a prefix.

\paragraph{Scope.}
This paper does not claim that FLP is false, incomplete, or merely an artifact of naive state abstraction. FLP remains the classical impossibility result for deterministic consensus in a fully asynchronous system with one crash failure \cite{fischer-lynch-paterson-1985}. The claim here is interpretive and structural: FLP constrains guaranteed terminal collapse under its model. It does not say that communication cannot accumulate causal structure, nor that all protocols should be understood only from the terminal value abstraction. The protocol comparison is limited to representative fault-tolerant consensus and atomic-broadcast protocols whose safety is organized around explicit evidence carriers: quorums, certificates, slices, committees, stake-weighted chain evidence, checkpoint votes, sampling histories, estimators, or DAG structure.

\paragraph{Contributions.}
The paper makes four contributions.
\begin{itemize}[leftmargin=1.2em]
\item It gives a self-contained vocabulary for reading consensus protocols as maps from communication-induced evidence states to terminal outputs: evidence domain, communication carrier, legitimacy extraction, evidence horizon, collapse rule, and repair/deferral rule.
\item It separates legitimacy from collapse: quorum evidence, committee eligibility, checkpoint votes, sampling confidence, estimator safety, or DAG support says that evidence is decision-grade, while the collapse policy says when that evidence is sufficient to emit a value, log entry, block, or prefix.
\item It rereads the FLP--set agreement--topology lineage as a sequence of constraints on terminal decision carriers, without changing the underlying theorems.
\item It compares real consensus protocols in a shared normal form, with main case studies and appendix instantiations that name each carrier, operation, horizon, collapse rule, and protocol reading.
\end{itemize}

\paragraph{Non-contributions.}
The paper does not prove a new lower bound, derive the listed protocols from a single formal model, or claim that richer histories evade FLP. It also does not replace protocol-specific safety or liveness proofs, and it does not rank protocols by performance. Its contribution is a denotational specification framework for comparing consensus protocols by the communication evidence they retain, the legitimacy evidence they accept, the horizon they inspect, the projection they emit, and the repair or deferral rule they use when that horizon is insufficient.

\section{Consensus as Communication}

We separate two notions that are usually conflated.

\paragraph{Semantic use of communication.}
The word ``communication'' is semantic here, in the concurrency-theoretic sense: execution is interaction among components. It does not mean only explicit network sends. It names the relational structure by which evidence becomes available and constraining: causality, dependency, certification, observation, inclusion, equivocation, and justification. Messages are one way to expose that structure; the carrier is the interaction structure recorded by the execution.

The degenerate case in which all components already have the same fixed value, or in which validity does not depend on distributed information, is not the phenomenon studied here. The paper concerns consensus as a distributed computation: information is initially separated, and correctness depends on the interaction structure that makes some terminal observation safe.

\begin{definition}[Communication carrier]
Given an evidence domain \(\mathcal{H}\), a communication carrier
\(H_t \in \mathcal{H}\) is the evidence state of the concurrent execution up to
time or logical horizon \(t\). Depending on the protocol instance, the observable
distinctions may include labelled events, proposals, votes, messages, receives,
timeout evidence, certificates, justified messages, local views, or
DAG/blocklace structure. The relevant relations may include local order, message
delivery, inclusion, certification, ratification, justification, equivocation,
and anchor reachability.
\end{definition}

\begin{definition}[Terminal consensus output]
A terminal consensus output is an order-1 artifact such as a value, a log position, a block sequence, or a finalized prefix. It is obtained by applying a protocol-specific projection to a communication carrier.
\end{definition}

Standard consensus specifies properties of the terminal output. This paper asks how that output is produced from the communication carrier.

\subsection{Collapse}

\begin{definition}[Collapse rule]
A collapse rule is a partial map
\[
C : H_t \rightharpoonup P
\]
from an evidence state to an order-1 prefix or value carrier. The map is partial because a protocol may refuse to decide when the visible evidence is insufficient.
\end{definition}

Collapse is not the same as legitimacy. A quorum certificate, commit certificate, or super-ratification witness says that some evidence is decision-grade. Collapse says which distinctions in the evidence state stop mattering once the protocol emits a value or prefix.

\subsection{Evidence horizon}

Every concrete decision is made from a finite visible cut of the evidence state, even if the protocol as a whole maintains an unbounded growing carrier. We call this cut the evidence horizon.

\begin{definition}[Evidence horizon]
An evidence horizon is the portion of $H_t$ that a protocol consults before attempting collapse. It may be a view, round, phase sequence, quorum-certificate chain, wave, anchor distance, slot frontier, or timeout window.
\end{definition}

If the horizon is too short or too compressed, progress may be blocked or unsafe. Protocols then need a repair, deferral, or fallback mechanism: view change, timeout certificates, randomized leaders, indirect rules, undecided slots, or later anchors.

\subsection{Denotational reading of evidence}

The evidence domain \(\mathcal{H}\) is the space of possible order-2 carriers for a protocol. A particular \(H_t \in \mathcal{H}\) is the communication carrier of an execution prefix: the evidence state produced by component interaction up to horizon \(t\). The carrier may be described using protocol vocabulary--ballots, promises, certificates, DAG vertices, anchors, justified messages, or protocol states--but those names appear in the specification only for the distinctions they make meaningful. The generic collapse-policy form leaves \(\mathcal{H}\) abstract. Each protocol instance chooses the evidence domain on which its legitimacy, horizon, and collapse maps are defined.

\begin{definition}[Evidence domain]
An evidence domain for a protocol is a preordered set \((\mathcal{H},\sqsubseteq)\). Its elements are evidence states. The relation \(h \sqsubseteq h'\) means that \(h'\) preserves the evidence of \(h\) and may add further distinctions. The order is semantic: two descriptions may differ in irrelevant detail while determining the same evidence state.
\end{definition}

\begin{definition}[Semantic collapse]
Let \(\mathrm{Adm}(h)\) be the admissibility predicate for evidence states and let
\[
\mathsf{obs} : \mathcal{H} \rightharpoonup \mathcal{P}
\]
be the terminal observation. For \(h \in \mathcal{H}\), write
\[
\uparrow_{\mathrm{Adm}} h
=
\{h' \mid h \sqsubseteq h' \land \mathrm{Adm}(h')\}
\]
for the admissible future cone of \(h\). A prefix \(p\) is determined at \(h\) when some admissible refinement of \(h\) produces \(p\), and all admissible refinements of \(h\) that can produce an observation produce that same observation:
\[
\exists h_0 \in \uparrow_{\mathrm{Adm}} h.\;
\mathsf{obs}(h_0)\downarrow \land \mathsf{obs}(h_0)=p
\]
and
\[
\forall h_1,h_2 \in \uparrow_{\mathrm{Adm}} h.\;
\mathsf{obs}(h_1)\downarrow \land \mathsf{obs}(h_2)\downarrow
\Rightarrow
\mathsf{obs}(h_1)=\mathsf{obs}(h_2)=p .
\]
Collapse is valid when the emitted order-1 artifact is invariant under all admissible refinements of the current evidence.
\end{definition}

\paragraph{Meaning before mechanism.}
Under this reading, quorum certificates, locks, anchors, committee credentials, sampling thresholds, and safety oracles are not the meaning of consensus. They are protocol-specific witnesses that a terminal observation has become invariant over the relevant admissible future cone. A protocol may establish this invariant through messages, votes, DAG reachability, or estimator analysis, but the semantic question is the same: have all remaining admissible refinements lost the ability to change the exposed value or prefix?

\paragraph{Collapse debt.}
We call this situation \emph{collapse debt}: the protocol tries to obtain an order-1 decision from an evidence horizon whose admissible refinements can still change the terminal observation. The debt is paid by waiting, carrying additional evidence, changing views, adding randomness, routing through fallback rules, or declaring a slot undecided.

Collapse debt measures the semantic gap between the evidence a protocol has chosen to inspect and the evidence needed to make its terminal observation invariant. A repair rule is therefore not merely an operational fallback; it is a way of preserving or recovering the communication distinctions that the first horizon could not yet lawfully discard.

For a concrete collapse attempt, debt is evaluated at the visible evidence \(v = R(L(H_t))\), not necessarily at the full carrier \(H_t\). A protocol may retain enough information in \(H_t\) to decide later while still refusing collapse from the current horizon.

This is a recurring link across protocol families. Many mechanisms presented as distinct innovations can be read as ways of managing collapse debt.

\section{Collapse-Policy Specification}

We describe protocols using the following normal form:
\[
H_t
\xrightarrow{\;L\;}
E_t
\xrightarrow{\;R\;}
E_t|_R
\xrightarrow{\;C\;}
P .
\]

\begin{itemize}[leftmargin=1.2em]
\item \(H_t \in \mathcal{H}\) is the current communication carrier: the order-2 evidence state visible at the chosen time or horizon.
\item $L$ is the legitimacy extraction that marks evidence as decision-grade.
\item $E_t$ is the admissible evidence extracted from the history.
\item $R$ is the evidence horizon: the bounded region the protocol is willing to inspect before output.
\item $E_t|_R$ is the visible admissible evidence inside the horizon.
\item $C$ is the collapse rule that emits a prefix/value only when the visible evidence determines an observation, or refuses to decide.
\item $P$ is the order-1 terminal artifact: value, log, block prefix, or ordered slot prefix.
\end{itemize}

The legitimacy extraction belongs to $L$. For crash-fault protocols this is usually majority intersection; for Byzantine quorum protocols it is usually the $2f+1$ supermajority rule; in weighted or flexible variants it is the required intersection relation among the relevant weighted or phase-specific quorums. In other families, legitimacy may come from quorum slices, VRF committee credentials, stable-prefix assumptions, checkpoint votes, sampling confidence, estimator safety, or DAG support. None of these is a collapse rule. Collapse belongs to $R$ and $C$: how long the protocol waits, what evidence remains visible, and what is projected away when output is produced.

\subsection{Protocol vocabulary dictionary}

The normal form does not erase protocol terminology; it maps protocol terms to semantic roles. The table is a guide rather than a strict partition: blocks may be carrier events or output artifacts, and certificates, leaders, or anchors may also shape the horizon or collapse rule.

\begin{center}
\small
\renewcommand{\arraystretch}{0.95}
\begin{tabular}{p{0.24\linewidth}p{0.69\linewidth}}
\textbf{Semantic role} & \textbf{Protocol terms} \\
\hline
evidence state \((H)\)
  & votes, accepts, proposals, blocks, ratifications \\
legitimacy extraction \((L)\)
  & quorum intersection, threshold checks, credentials, confidence, equivocation filters, safety predicates \\
admissible evidence \((E)\)
  & quorums, certificates, accepted values, ratifications, confidence counters, support/skip evidence \\
visibility horizon \((R)\)
  & leaders, proposers, primaries, anchors, relays, rounds, terms, views, waves, slots, QC chains \\
collapse and repair \((C)\)
  & timeouts, view changes, fallback rules, random leaders, indirect rules \\
terminal output \((P)\)
  & values, log entries, blocks, finalized prefixes
\end{tabular}
\end{center}

\subsection{Specification form}

The specification below makes the normal form explicit for the consensus protocols considered here, both in the main case studies and in the appendix catalog. It does not restate each protocol proof. It gives a small specification language in which each protocol becomes an instance: first the carriers and operations, then the mechanisms that realize them.

\noindent The examples are representative published consensus or atomic-broadcast protocol families. The main text focuses on Lamport causal ordering, Byzantine Generals, Paxos, HotStuff, CBC Casper, Cordial Miners, and Mysticeti-C. The appendix gives additional instantiations for HoneyBadgerBFT and randomized asynchronous agreement, Raft, PBFT, Stellar, Tendermint, Streamlet, Algorand, Ouroboros, GRANDPA, Casper FFG, Snowman/Avalanche, Jolteon/Ditto, DAG-Rider, Hashgraph, Aleph, and Narwhal/Tusk/Bullshark.

\noindent The method is spec-first. A protocol is introduced as a collapse-policy specification over named carriers: the carriers are named sorts, the operations are \(L\), \(R\), and \(C\), and the laws say when the specified meaning is preserved. The protocol vocabulary names the protocol-specific evidence domain and the witnesses that satisfy those laws.

\begin{definition}[Collapse-policy specification]
A collapse-policy specification for a protocol family \(\Pi\) consists of named carriers
\begin{specdisplay}
\mathcal{H}_\Pi &:& \mathsf{Type} \\
\mathcal{E}_\Pi &:& \mathsf{Type} \\
\mathcal{V}_\Pi &:& \mathsf{Type} \\
\mathcal{P}_\Pi &:& \mathsf{Type}
\end{specdisplay}
and operations
\begin{specdisplay}
L_\Pi &:& \mathcal{H}_\Pi \to \mathcal{E}_\Pi \\
R_\Pi &:& \mathcal{E}_\Pi \to \mathcal{V}_\Pi \\
C_\Pi &:& \mathcal{V}_\Pi \rightharpoonup \mathcal{P}_\Pi .
\end{specdisplay}
Here \(\mathcal{H}_\Pi\) is the protocol-specific evidence domain, \(\mathcal{E}_\Pi\) is decision-grade evidence, \(\mathcal{V}_\Pi\) is the visible evidence inside the chosen horizon, and \(\mathcal{P}_\Pi\) is the terminal output carrier. The induced decision map is
\[
D_\Pi \;\mathrel{:=}\; C_\Pi \circ R_\Pi \circ L_\Pi
\qquad
D_\Pi : \mathcal{H}_\Pi \rightharpoonup \mathcal{P}_\Pi .
\]
When \(C_\Pi\) is undefined, the protocol does not have a valid collapse from the current horizon. It must either preserve the carrier for later evidence, repair the horizon, or defer output.
\end{definition}

\begin{definition}[Realization law]
A concrete realization \(I\) of a collapse-policy specification \(\Pi\) supplies carriers
\(\mathcal{H}^I_\Pi,\mathcal{E}^I_\Pi,\mathcal{V}^I_\Pi,\mathcal{P}^I_\Pi\), concrete operations
\(L^I_\Pi,R^I_\Pi,C^I_\Pi\), and denotations
\[
\denote{-}_\mathcal{H} :
\mathcal{H}^I_\Pi \to \mathcal{H}_\Pi,\quad
\denote{-}_\mathcal{E} :
\mathcal{E}^I_\Pi \to \mathcal{E}_\Pi,\quad
\denote{-}_\mathcal{V} :
\mathcal{V}^I_\Pi \to \mathcal{V}_\Pi,\quad
\denote{-}_\mathcal{P} :
\mathcal{P}^I_\Pi \to \mathcal{P}_\Pi .
\]
Correctness means that the realization preserves the specified meaning:
\[
\begin{aligned}
\denote{L^I_\Pi(h)}_\mathcal{E}
&\equiv
L_\Pi(\denote{h}_\mathcal{H}), \\
\denote{R^I_\Pi(e)}_\mathcal{V}
&\equiv
R_\Pi(\denote{e}_\mathcal{E}), \\
\denote{C^I_\Pi(v)}_\mathcal{P}
&\equiv
C_\Pi(\denote{v}_\mathcal{V})
\quad\text{when } C^I_\Pi(v) \text{ is defined}.
\end{aligned}
\]
Consequently,
\[
\denote{D^I_\Pi(h)}_\mathcal{P}
\equiv
D_\Pi(\denote{h}_\mathcal{H})
\]
whenever the realization emits an output. This is the denotational obligation: concrete structures may vary, but their observable collapse must preserve the specified meaning.
\end{definition}

\section{FLP as Order-Collapse Constraint}

The FLP theorem states that deterministic consensus cannot guarantee termination in a fully asynchronous message-passing system with even one crash failure \cite{fischer-lynch-paterson-1985}. It is a theorem about the classical terminal task.

The order vocabulary does not change the theorem. It makes explicit which semantic operation the theorem constrains. FLP is about the impossibility of guaranteeing an order-1 terminal decision from an asynchronous execution carrier under a particular adversary model: deterministic processes, reliable asynchronous message passing, and one possible crash failure.

\paragraph{Classical structure.}
The FLP proof reasons over global configurations: process states, message buffers, and the deterministic protocol state induced by prior steps. A configuration is \emph{0-valent} if every admissible continuation decides 0, \emph{1-valent} if every admissible continuation decides 1, and \emph{bivalent} if both decisions remain possible under different admissible continuations. The proof shows that there is an initial bivalent configuration and that, from a bivalent configuration, the asynchronous adversary can choose a next event that keeps the execution bivalent.

This is not a claim that FLP uses an impoverished state model. FLP configurations are already rich enough to include process state and pending-message information. The collapse reading instead identifies the operation constrained by the theorem: a deterministic protocol cannot guarantee that every admissible execution eventually reaches a region where the terminal decision map is determined.

\paragraph{Translation to the normal form.}
In this notation, the growing communication carrier is \(H_t\). The protocol extracts decision-grade evidence \(E_t\), restricts attention to a horizon \(E_t|_R\), and applies a collapse rule
\[
C : E_t|_R \rightharpoonup P
\]
where, for binary consensus, \(P=\{0,1\}\). A univalent configuration is one where the remaining admissible histories already lie in one fiber of this collapse: all completions compatible with the current evidence lead to the same decision. A bivalent configuration is one where the current history and visible evidence are still compatible with multiple fibers of \(C\).

\paragraph{What the adversary preserves.}
The FLP adversary does not need to corrupt messages or break safety. It exploits asynchrony: delay and crash are indistinguishable to finite local views. By choosing which pending message or process step becomes visible next, the adversary keeps the current \(H_t\) inside a region where both decision fibers remain reachable. The current carrier has not semantically closed over a single order-1 decision.

\paragraph{Concrete reading.}
Suppose a process is about to take a step that would make decision 0 inevitable. FLP's commutativity and indistinguishability argument shows that, in the asynchronous crash model, there is another admissible scheduling of independent or delayed events that cannot be distinguished locally but preserves the possibility of decision 1. The protocol state may grow, and messages may be delivered, but the evidence visible inside the current horizon still does not determine a unique collapse. The system can keep accumulating communication-induced evidence without reaching a point where every fair admissible continuation must project to the same terminal value.

\paragraph{Order-1 progress condition.}
This is why FLP is central to the thesis. The safety condition says that if collapse happens, all correct processes must collapse compatibly. The termination condition says collapse must eventually happen. FLP says that, under full asynchrony and one crash failure, deterministic protocols cannot guarantee both for the order-1 consensus output. The obstruction is not communication itself; it is guaranteed terminal collapse.

\paragraph{Common overreading.}
The overreading is to treat FLP as saying that distributed coordination itself cannot progress under asynchrony. That is too broad. Communication can still accumulate causal structure. Processes can still exchange evidence, form local views, build partial orders, and preserve alternatives. What FLP rules out is deterministic guaranteed collapse to a terminal decision under the asynchronous crash assumptions.

\paragraph{Why this matters.}
Once consensus is viewed only from the terminal output, all pre-decision communication structure becomes protocol plumbing. Once the communication carrier is made explicit, FLP appears as a limit on guaranteed collapse, not as a denial that the order-2 carrier can grow. This is the key separation: evidence growth belongs to \(H_t\), while decision belongs to \(C\). FLP constrains the guarantee that \(C\) must become defined, not the existence of meaningful structure in \(H_t\).

\paragraph{Ways protocols escape the FLP setting.}
Protocols obtain progress by changing the problem setting or the collapse policy: adding partial synchrony \cite{dwork-lynch-stockmeyer-1988}, adding failure detectors \cite{chandra-toueg-1996}, using randomization \cite{keidar-dagrider-2021}, relying on stronger timing/eventual-leader behavior, widening the output carrier, or deferring collapse until richer evidence is visible. These are not violations of FLP. They change the conditions under which \(C\) is allowed, expected, or required to produce an output.

\section{Collapse Ladder in Consensus Theory}

With FLP stated in normal form, the rest of the impossibility lineage can be read as a progressive exposure of collapse structure.

\paragraph{FLP: binary collapse.}
FLP is the binary base case. The decision set has width one: all correct processes must collapse to the same value, even though the value domain has two alternatives. The adversary preserves bivalence, so the collapse rule \(C\) cannot be guaranteed to become defined in every admissible execution \cite{fischer-lynch-paterson-1985}.

\paragraph{Chaudhuri: collapse width.}
Set agreement makes the output width explicit. In $k$-set agreement, processes decide proposed values, but at most $k$ distinct values may be decided. Consensus is the special case $k=1$. Chaudhuri's formulation therefore exposes $k$ as the allowed residual ambiguity after collapse: more output width can absorb more unresolved alternatives. The possibility side already has this shape: there is a simple $(k-1)$-resilient protocol for $k$-set agreement in a totally asynchronous system \cite{chaudhuri-set-consensus-1990}.

\paragraph{Borowsky--Gafni: resilient simulation.}
Borowsky and Gafni generalize the FLP pattern by showing that the obstruction scales with resilience \cite{borowsky-gafni-1993}. In particular, the matching lower-bound reading is that $k$-set consensus cannot be guaranteed when the protocol must tolerate $k$ crash failures. The useful reading here is not just ``more impossibility.'' It is that crash resilience gives the adversary scheduling power to preserve a space of alternatives too large for the requested output carrier. FLP is bivalence; the generalized picture is polyvalence.

\paragraph{Herlihy--Shavit: topological factorization.}
Topological distributed computing gives the geometric form of the same claim. A task is solvable when the protocol complex, which records executions and local views, admits a suitable simplicial map into the output/task complex \cite{herlihy-shavit-topology}. Saks and Zaharoglou's public-knowledge proof gives the parallel topological impossibility for wait-free $k$-set agreement \cite{saks-zaharoglou-1993}. Impossibility means no structure-preserving collapse exists from the history/view carrier to the decision carrier.

\paragraph{Interpretation.}
This ladder suggests a semantic diagnosis. Once the problem is framed as agreement on an order-1 artifact, protocol research naturally studies mechanisms for making that terminal artifact safe and live under adversarial uncertainty. Leaders, rounds, views, quorum certificates, locks, timeouts, waves, anchors, randomization, failure detectors, direct rules, indirect rules, and undecided slots can be read as ways of managing the boundary between a rich communication carrier and a narrower decision carrier.

\section{From Impossibility to Protocol Mechanisms}

The ladder above should not be read as a formal derivation of Paxos, PBFT, HotStuff, or DAG protocols. The models differ: FLP and Borowsky--Gafni are crash-failure impossibility results; topological computability is usually stated for wait-free read/write shared memory and colorless tasks; later BFT protocols handle Byzantine faults, partial synchrony, randomness, authentication, threshold evidence, and replicated logs.

The connection is semantic rather than model-identical. The impossibility lineage identifies the obstruction to forcing collapse under adversarial uncertainty. Practical protocols then choose how to change the conditions around that obstruction:
\begin{itemize}[leftmargin=1.2em]
\item \textbf{add timing:} partial synchrony and leader responsiveness make some evidence eventually timely;
\item \textbf{add oracles:} failure detectors and view-change logic introduce information not present in pure asynchrony;
\item \textbf{add randomness:} randomized protocols prevent the adversary from deterministically preserving the same ambiguity forever;
\item \textbf{add authentication and quorum evidence:} certificates make local claims portable across processes and views;
\item \textbf{delay collapse:} DAG and wave protocols retain more communication-derived evidence before projecting to order;
\item \textbf{refuse collapse:} undecided slots, failed waves, and fallback paths avoid emitting a prefix across insufficient evidence.
\end{itemize}

Thus the protocol comparison below is not a claim that every mechanism is an instance of one theorem. It is a normal-form comparison of how protocols manage the same semantic pressure: an order-2 communication carrier must eventually be interpreted as an order-1 value, log, or prefix.

\section{Protocol Specifications and Readings}

This section applies the normal form to representative protocols. The point is not that all protocols are identical. The point is that their apparent differences can be placed in the same comparison frame: legitimacy, horizon, collapse, and repair.

The main text does not attempt to survey every consensus protocol in equal detail. Instead, it develops the denotational argument through cases that expose distinct semantic roles: causal projection, carrier enrichment, quorum recovery, certificate compression, admissible-future safety, and DAG collapse debt. The appendix then instantiates the same normal form for a broader set of protocols. Those appendix entries are not secondary in protocol importance; they are secondary in the proof strategy. They serve as coverage checks once the semantic argument has been established.

\begin{center}
\small
\renewcommand{\arraystretch}{1.05}
\begin{tabular}{@{}p{0.18\linewidth}p{0.25\linewidth}p{0.25\linewidth}p{0.24\linewidth}@{}}
\textbf{Case} & \textbf{Evidence domain} & \textbf{Semantic role} & \textbf{Collapse witness} \\
\hline
Lamport ordering
  & events, messages, logical clocks
  & causal carrier before total order
  & total-order extension of happened-before \\
Byzantine Generals
  & oral or signed report histories
  & carrier enrichment under equivocation
  & loyal-consistent interpretation of reported values \\
Paxos
  & ballots, promises, accepted values
  & crash-fault quorum recovery
  & majority evidence plus highest accepted ballot \\
HotStuff
  & proposal tree, votes, QCs
  & Byzantine certificate compression
  & certified-chain ancestry rule \\
CBC Casper
  & justified message sets
  & admissible-future safety
  & estimator invariance over future states \\
Cordial Miners
  & blocklace waves and ratifications
  & backward evidence from final leaders
  & final leader and ordering function \(\tau\) \\
Mysticeti-C
  & slot DAG, anchors, support/skip evidence
  & explicit collapse debt at a slot frontier
  & direct or indirect slot classification \\
\end{tabular}
\end{center}

The section is organized spec-first. For each main case, the named carriers and maps state the semantic object; the protocol reading then explains how the protocol mechanisms realize that object. The comparative notes preserve the original protocol-by-protocol reading while making the specification the primary artifact.

In the protocol specifications below, the displayed carriers are protocol-specific evidence domains. The purpose of each main case is to identify the witness by which the protocol establishes invariance of the terminal observation. Concrete realizations enter only through the realization law above: they must denote these specified carriers and preserve the specified collapse.

\subsection{Reference baseline}

The reference baseline keeps the communication carrier as the primary semantic object. A finalization rule commits causally closed subhistories rather than immediately projecting them to a single prefix. A prefix output may still be produced, but it is explicitly a projection from finalized history.

This baseline is not presented as a practical protocol. It is the reference object against which information loss is measured:
\begin{itemize}[leftmargin=1.2em]
\item \textbf{carrier:} labelled event DAG with causal and evidential edges;
\item \textbf{legitimacy:} threshold-valid evidence inside the DAG;
\item \textbf{horizon:} delayed until the relevant causal subhistory is visible;
\item \textbf{collapse:} optional projection to a prefix after order-2 finalization;
\item \textbf{loss:} only the loss introduced by the explicit projection, not by finalization itself.
\end{itemize}

\subsection{Lamport causal ordering}

The pre-consensus foundation is Lamport's happened-before relation and logical clocks \cite{lamport-time-clocks-1978}. That paper is not a consensus protocol in the later Paxos/Raft/PBFT sense, but it is the starting point for this communication-history view. It identifies distributed execution as a partially ordered set of events and then shows how logical clocks can produce a total order extension of that partial order.

In the normal form:
\begin{itemize}[leftmargin=1.2em]
\item \textbf{carrier:} distributed events, process-local order, and message send/receive causality;
\item \textbf{legitimacy:} happened-before evidence induced by local order and communication;
\item \textbf{horizon:} events visible to a process through received messages and logical-clock updates;
\item \textbf{collapse:} logical timestamps plus tie-breaking extend a causal partial order into a total order;
\item \textbf{discarded distinctions:} concurrency is hidden when incomparable events are placed in an arbitrary total order;
\item \textbf{repair mechanism:} none at the consensus level; this is ordering machinery, not fault-tolerant agreement.
\end{itemize}

This is the original carrier/projection pattern for distributed ordering. A communication-induced partial order is the primary object, and a total order is a projection used when applications require a single sequence. Later consensus protocols add faults, quorums, certificates, leaders, and finality, but they still manage the same boundary: when can communication-induced evidence be collapsed into a terminal order or prefix?

\subsection{Byzantine Generals as local-history collapse}

The Byzantine Generals Problem is the Byzantine base case of the collapse view \cite{lamport-shostak-pease-1982}. Lamport, Shostak, and Pease do not begin with a blockchain-style finalized prefix or even with a replicated log. They begin with a more primitive question: how can loyal participants obtain compatible information when faulty participants may send conflicting reports to different places?

The paper first frames the generals' decision as a function of values \(v(1),\ldots,v(n)\), where \(v(i)\) is the value communicated by general \(i\). The difficulty is that a traitorous general may send different values to different loyal generals. Therefore, before any robust decision function such as majority can be applied, loyal generals must agree on which value of \(v(i)\) is to be used. This is precisely a local-history problem: different loyal participants may hold different communication histories that are locally plausible but cannot all be collapsed to the same decision.

In the normal form:
\begin{itemize}[leftmargin=1.2em]
\item \textbf{carrier:} oral reports or signed reports about who said what;
\item \textbf{legitimacy:} oral relay structure in the unsigned model, transferable signature evidence in the signed model;
\item \textbf{horizon:} recursive message depth for oral messages, or signature paths for signed messages;
\item \textbf{collapse:} a choice rule, such as majority, applied only after each \(v(i)\) has a loyal-consistent interpretation;
\item \textbf{discarded distinctions:} alternate reports and equivocations not retained by the final selected value;
\item \textbf{repair mechanism:} deeper oral recursion, or carrier enrichment by unforgeable signatures.
\end{itemize}

This makes the role of signatures especially clear. A signed message is not merely an authentication optimization. It changes the semantic carrier: equivocation becomes portable evidence. In the oral-message model, a report about a report may need additional relay depth before loyal participants can interpret it consistently. In the signed-message model, conflicting signed statements can be carried forward as first-class evidence.

Thus the Byzantine Generals Problem is not only the origin of Byzantine fault tolerance; it is an early instance of the pattern. Byzantine faults attack the communication carrier itself. The protocol must preserve or enrich enough of that carrier before collapse to a value is safe.

\subsection{Paxos and Multi-Paxos}

Consider one Paxos instance. A proposer enters ballot \(b\) and asks acceptors for promises. The carrier contains proposals, promises, accepted ballot/value pairs, and acceptor state. The legitimacy backbone is majority intersection: once a majority of acceptors has responded, later majorities intersect with earlier ones, so accepted evidence can be recovered.

The horizon is the phase-1 response set. If that response set reveals that some acceptor has already accepted a value, the proposer cannot safely ignore it and propose something arbitrary. It must recover the value with the highest accepted ballot and then seek majority accepts for that value. Multi-Paxos repeats this across slots under a stable leader, which makes the log case a repetition of the same recovery logic. The repair path is therefore not "undo the history"; it is "lift the prior accepted evidence into the next ballot."

\paragraph{Protocol vocabulary.}
Proposer, ballot, promise, accepted value, acceptor state, majority.

\paragraph{Collapse-policy specification.}
\begin{specdisplay}
\mathcal{H}_{\mathrm{Paxos}} &:& \specdesc{proposals, ballots, promises,\\ accepted values, acceptor state} \\
\mathcal{E}_{\mathrm{Paxos}} &:& \specdesc{accepted ballot/value evidence\\ from acceptors} \\
\mathcal{V}_{\mathrm{Paxos}} &:& \specdesc{phase-1 evidence visible in\\ ballot $b$, followed by phase-2 accepts} \\
\mathcal{P}_{\mathrm{Paxos}} &:& \text{chosen value for slot } s \\[0.4em]
L_{\mathrm{Paxos}}(h) &\mathrel{:=}&
\{(a,b',x) \mid a \text{ accepted value } x \text{ in ballot } b' \text{ in } h\} \\
R_{\mathrm{Paxos},b}(e) &\mathrel{:=}&
\{(a,b',x) \in e \mid a \text{ promised ballot } b\} \\
S_{\mathrm{Paxos},x_0}(v) &\mathrel{:=}&
\begin{cases}
x & \text{if } (a,b',x) \in v \text{ has maximal } b', \\
x_0 & \text{if } v \text{ contains no accepted value.}
\end{cases}
\\
C_{\mathrm{Paxos}}(v) &\mathrel{:=}&
\text{the value accepted by a majority in phase 2, when such a majority exists.}
\end{specdisplay}

\paragraph{Protocol reading.}
Promises and accepted values carry the prior evidence, and the phase-1 response set is the horizon for choosing a safe proposal \(S_{\mathrm{Paxos}}\). Majority accepts in phase 2 implement the collapse \(C_{\mathrm{Paxos}}\). Multi-Paxos repeats the same specification across slots under a stable leader.

\paragraph{Denotational reading.}
For one Paxos slot, \(h \sqsubseteq h'\) means that \(h'\) extends the ballot history while preserving acceptor promises and accepted-value evidence already present in \(h\). The admissibility predicate \(\mathrm{Adm}\) restricts refinements to executions that respect the Paxos acceptor rules: an acceptor does not accept below a promised ballot, and majorities intersect. The terminal observation \(\mathsf{obs}(h)\) is the value chosen for the slot, when such a value is forced by majority accept evidence. Phase 1 is therefore a meaning test, not the final collapse: the proposer asks whether the current evidence state already lies in a future cone where some prior value constrains all safe proposals. If so, the highest accepted value is the only value whose observation can remain invariant under admissible extension; collapse occurs only when phase 2 obtains a majority of accepts for that value.

\paragraph{Comparative reading.}
Paxos turns majority quorum intersection into a crash-fault consensus mechanism \cite{lamport-paxos-1998,lamport-paxos-simple-2001}. In the normal form:
\begin{itemize}[leftmargin=1.2em]
\item \textbf{carrier:} proposals, ballots, promises, accepted values, acceptor state, and learner observations;
\item \textbf{legitimacy:} intersecting majorities of acceptors;
\item \textbf{horizon:} ballot phase and accepted-value evidence visible to a proposer;
\item \textbf{collapse:} a majority of accepts chooses a value for an instance; Multi-Paxos repeats this over log slots under a stable leader;
\item \textbf{discarded distinctions:} losing proposals, alternate proposer races, and timing histories not reflected in the chosen value;
\item \textbf{repair mechanism:} higher ballots recover prior accepted evidence before proposing a safe value.
\end{itemize}

From the collapse view, Paxos is already a collapse-debt protocol. A proposer cannot safely choose an arbitrary value after earlier ballots may have made progress. It must first ask acceptors what evidence already exists, then collapse only through a value compatible with that evidence.

\subsection{HotStuff}

HotStuff's concrete object is a proposal tree under rotating leaders. The carrier contains proposal edges, votes, quorum certificates, locks, highest-QC information, and timeout evidence in later variants. The legitimacy relation is a QC, which makes support for a block portable across validators and views. The horizon is the certified chain, classically the chain depth used by the three-chain commit rule.

Collapse commits an ancestor of the certified chain and extends the ordered prefix. The QC chain compresses several rounds of evidence into a structure that can be carried and reused across view changes. If a leader fails to produce the needed chain, the protocol repairs by rotating leaders, propagating the highest QC, and using lock and timeout evidence to keep safety while preserving liveness.

\paragraph{Protocol vocabulary.}
Proposal tree, votes, quorum certificate, lock, highest QC, timeout, view change, leader rotation.

\paragraph{Collapse-policy specification.}
\begin{specdisplay}
\mathcal{H}_{\mathrm{HotStuff}} &:& \specdesc{proposal tree, votes, QCs, locks, highest-QC state,\\ timeout evidence} \\
\mathcal{E}_{\mathrm{HotStuff}} &:& \specdesc{certified proposal-chain evidence} \\
\mathcal{V}_{\mathrm{HotStuff}} &:& \specdesc{certified chain visible in the current view} \\
\mathcal{P}_{\mathrm{HotStuff}} &:& \text{ordered prefix} \\[0.4em]
L_{\mathrm{HotStuff}}(h) &\mathrel{:=}&
\{(b,q) \mid q \text{ is a quorum certificate for block } b \text{ in } h\} \\
R_{\mathrm{HotStuff},v}(e) &\mathrel{:=}&
\{(b,q) \in e \mid (b,q) \text{ is carried into view } v\} \\
C_{\mathrm{HotStuff}}(u) &\mathrel{:=}&
\{a \mid u \text{ contains a certified three-chain committing ancestor } a\}.
\end{specdisplay}

\paragraph{Protocol reading.}
The proposal tree is the carrier, the QC is the legitimacy witness, the certified chain is the horizon, and the three-chain rule collapses that chain into an ordered prefix while leader rotation preserves safety across failures.

\paragraph{Denotational reading.}
For HotStuff, \(h \sqsubseteq h'\) means that \(h'\) extends the proposal tree with additional votes, certificates, timeout evidence, and descendants while preserving certified ancestry already visible in \(h\). The admissibility predicate \(\mathrm{Adm}\) enforces validator voting rules, quorum thresholds, and lock-compatible extensions. The observation \(\mathsf{obs}(h)\) is the committed prefix determined by certified ancestry. A quorum certificate is not itself the meaning of finality; it is a portable witness that a block has entered the evidence state strongly enough to constrain later admissible refinements. The three-chain rule is the point at which further lock-compatible refinements can no longer change the committed ancestor observation.

\paragraph{Comparative reading.}
HotStuff keeps the Byzantine quorum threshold but compresses evidence into quorum-certificate chains under rotating leaders \cite{yin-hotstuff-2019}. In the normal form:
\begin{itemize}[leftmargin=1.2em]
\item \textbf{carrier:} proposal tree, votes, quorum certificates, locks, view changes, timeout evidence in later variants;
\item \textbf{legitimacy:} a quorum certificate for a proposed block;
\item \textbf{horizon:} a chain of certified proposals, classically the three-chain commit rule;
\item \textbf{collapse:} the QC chain commits an ancestor block and extends the ordered prefix;
\item \textbf{discarded distinctions:} uncertified branches, alternate leader proposals, and causal details compressed into the QC chain;
\item \textbf{repair mechanism:} highest-QC propagation, leader rotation, view synchronization, and timeout certificates in practical variants.
\end{itemize}

HotStuff keeps legitimacy in quorum certificates and changes the shape of the carrier. Multi-phase evidence becomes a certified chain, so the commit rule reads ancestry in that chain rather than a standalone phase transcript.

\paragraph{Two-chain descendants.}
The three-chain rule is the classical HotStuff point in this design space, not a claim that three certified links are semantically primitive. From the order perspective, the later two-chain variants expose a tradeoff that is easy to miss in a phase-count presentation. Jolteon uses a two-chain commit rule by accepting a quadratic view-change path, reducing the steady-state latency of standard three-chain HotStuff \cite{gelashvili-jolteon-ditto-2022}. HotStuff-2 shows the same lesson in a sharper form: two phases can be enough while retaining the desired HotStuff properties when the synchronization and view-change evidence is organized differently \cite{malkhi-hotstuff2-2023}. If the minimum commit horizon is two certified order steps, then a three-chain rule is not a semantic necessity; it is a choice to keep more of the safety argument inside the visible proposal chain. Reducing the chain horizon to two requires the missing order constraints to be carried elsewhere, in the synchronization or view-change evidence. Thus the design parameter is not simply ``two phases versus three phases'', but the distribution of order evidence between the proposal-chain carrier and the repair carrier.

\subsection{CBC Casper}

CBC Casper starts from a general proof shape. A protocol state \(\sigma\) is a set of messages. A message carries an estimate, a sender, and a justification. The justification is itself a set of earlier messages, so protocol states carry their own dependency structure. The estimator \(E\) maps a protocol state to a consensus estimate: in the blockchain instance, the GHOST fork choice.

Faults are represented by equivocation evidence. A validator equivocates when it has two messages in the dependency closure of a state such that neither message depends on the other. The admissible protocol states \(\Sigma_t\) are those whose evidenced fault weight is at most \(t\). A state transition is set extension inside \(\Sigma_t\). Finality is then estimate safety: an estimate is safe when every admissible future state preserves it.

\paragraph{Protocol vocabulary.}
Protocol state, message, estimate, sender, justification, dependency, latest message, estimator, GHOST, equivocation, fault weight, safety oracle, safe estimate.

\paragraph{Collapse-policy specification.}
\begin{specdisplay}
\mathcal{H}_{\mathrm{CBC}} &:& \specdesc{sets of valid messages with justification dependencies} \\
\mathcal{E}_{\mathrm{CBC}} &:& \specdesc{latest-message, estimator, equivocation,\\ fault-weight evidence} \\
\mathcal{V}_{\mathrm{CBC}} &:& \specdesc{candidate estimate with admissible-future analysis} \\
\mathcal{P}_{\mathrm{CBC}} &:& \text{safe estimate or finalized block} \\[0.4em]
L_{\mathrm{CBC}}(\sigma) &\mathrel{:=}&
\left\langle \sigma, E(\sigma), \mathrm{latest}(\sigma), \mathrm{Eq}(\sigma) \right\rangle \\
\mathrm{Eq}(\sigma) &\mathrel{:=}&
\{(m_1,m_2) \in D(\sigma)^2 \mid
\mathrm{sender}(m_1)=\mathrm{sender}(m_2),\\
&&\qquad\qquad
m_1 \not\prec m_2,\; m_2 \not\prec m_1\} \\
\Sigma_t &\mathrel{:=}&
\{\sigma \subseteq M \mid F(\sigma) \leq t\} \\
\sigma \to_t \sigma' &\mathrel{:=}&
\sigma,\sigma' \in \Sigma_t \land \sigma \subseteq \sigma' \\
\mathrm{Safe}_t(e,\sigma) &\mathrel{:=}&
\forall \sigma' \in \Sigma_t.\; \sigma \to_t \sigma' \Rightarrow e \equiv E(\sigma') \\
R_{\mathrm{CBC},e}(l) &\mathrel{:=}&
\text{the safety-oracle view of } \mathrm{Safe}_t(e,\sigma) \text{ where } \sigma \text{ is carried by } l \\
C_{\mathrm{CBC}}(u) &\mathrel{:=}&
\begin{cases}
e & \text{if } u \text{ certifies } \mathrm{Safe}_t(e,\sigma),\\
\text{undefined} & \text{otherwise.}
\end{cases}
\end{specdisplay}

For the blockchain instance, \(e \equiv E(\sigma')\) means that the candidate block \(e\) remains in the chain selected by the estimator \(E(\sigma')\). Thus finality is not equality with one current fork choice; it is invariance of the relevant prefix across all admissible future fork choices.

\paragraph{Protocol reading.}
Messages and justifications are the carrier, the estimator supplies the current consensus proposition, equivocation evidence bounds admissible futures, and the safety oracle is the horizon test. CBC collapses only when the candidate estimate is invariant over those futures. If the oracle cannot certify safety, the protocol must keep the carrier open by receiving or producing more justified messages; the core CBC safety construction does not provide liveness by itself.

\paragraph{Comparative reading.}
Correct-by-Construction Casper, in Vlad Zamfir's Casper the Friendly Ghost presentation, starts from protocol states, estimates, and a safety theorem rather than from a leader, phase sequence, or commit certificate \cite{zamfir-casper-tfg-2017}. A protocol state is a set of valid messages. Each message carries an estimate, a sender, and a justification: the prior messages that make the new estimate meaningful. The estimator maps a protocol state to the current consensus proposition: a bit in the binary version, or a GHOST fork-choice block in the blockchain version.

The collapse point is not the estimator by itself. The estimator gives the current estimate. Finality requires estimate safety: an estimate is safe in state \(\sigma\) when every admissible future state reachable from \(\sigma\) keeps the estimator compatible with that estimate. Equivocation evidence determines which future states are admissible under the fault threshold \(t\). In the blockchain instance, a block is safe when it remains in the fork choice for every future protocol state with fault weight at most \(t\).

In the normal form:
\begin{itemize}[leftmargin=1.2em]
\item \textbf{carrier:} sets of valid messages or blocks, each with estimate, sender, and justification dependencies;
\item \textbf{legitimacy:} estimator-valid messages, latest-message weight, and equivocation/fault evidence below threshold \(t\);
\item \textbf{horizon:} the local protocol state \(\sigma\), plus the safety oracle's analysis of admissible future extensions of \(\sigma\);
\item \textbf{collapse:} decide an estimate or finalize a block only when it is safe across all admissible futures;
\item \textbf{discarded distinctions:} future branches or message extensions that can no longer change the safe estimate;
\item \textbf{repair mechanism:} if the safety oracle cannot certify safety, continue accumulating justified messages or equivocation evidence; the core safety construction does not by itself specify a liveness strategy.
\end{itemize}

CBC Casper gives the cleanest set-theoretic presentation of the order-theoretic reading, because protocol-state extension is explicit set inclusion. But the set is not the semantic source of finality. It is a representation of communication evidence: messages carry justifications, justifications induce dependency order, and adding messages refines the state. The primitive object is still the growing communication structure, while set inclusion is the chosen mathematical presentation of its refinement order.

For this reason, CBC is used here as evidence for the comparison method, not as the comparison method itself. Taking message sets as primitive would hide the broader point: different protocols present the same continuation question through different carriers of communication order.

Finality is therefore not a special chain-depth pattern but invariance of the estimator over all admissible future extensions. In this sense CBC makes explicit what chain protocols encode operationally: collapse is justified when further admissible communication can no longer change the observation. Once protocol states, estimators, and admissible futures are named, finality can be stated as estimate safety rather than as a protocol-specific commit gadget. The collapse-policy view then asks the same questions across other carriers: why this carrier is needed, how the horizon is chosen, and what order distinctions are lost when it projects to a lower-order output.

\subsection{Cordial Miners}

Cordial Miners uses a blocklace to support dissemination, equivocation exclusion, and ordering. The carrier contains blocks, acknowledgment or ratification edges, equivocation evidence, and the partial order induced by the blocklace. The legitimacy relation is ratification and super-ratification by a supermajority of blocks. The horizon is the wave and the evidence for whether a final leader exists in that wave.

Collapse applies the deterministic ordering function \(\tau\), walking from final leaders through ratified structure to emit ordered fragments. If no final leader exists, the protocol defers ordering rather than forcing a prefix. Later blocklace evidence may create a final leader and let \(\tau\) settle earlier ambiguity.

\paragraph{Protocol vocabulary.}
Blocklace, acknowledgment, ratification, super-ratification, wave, final leader, ordering function \(\tau\).

\paragraph{Collapse-policy specification.}
\begin{specdisplay}
\mathcal{H}_{\mathrm{CM}} &:& \specdesc{blocklace blocks, acknowledgments, ratification edges,\\ equivocation evidence} \\
\mathcal{E}_{\mathrm{CM}} &:& \specdesc{ratified and super-ratified blocklace evidence} \\
\mathcal{V}_{\mathrm{CM}} &:& \specdesc{wave evidence with final-leader candidates} \\
\mathcal{P}_{\mathrm{CM}} &:& \text{ordered fragments} \\[0.4em]
L_{\mathrm{CM}}(h) &\mathrel{:=}&
\{b \in \mathrm{blocks}(h) \mid b \text{ is ratified or super-ratified in } h\} \\
R_{\mathrm{CM},w}(e) &\mathrel{:=}&
\{b \in e \mid b \text{ belongs to wave } w \text{ or is reachable from it}\} \\
C_{\mathrm{CM}}(u) &\mathrel{:=}&
\begin{cases}
\tau(\ell,u) & \text{if } \ell \text{ is a final leader visible in } u,\\
\text{undefined} & \text{otherwise.}
\end{cases}
\end{specdisplay}

\paragraph{Protocol reading.}
The blocklace is the carrier, ratification and super-ratification are the legitimacy witnesses, the wave and final-leader evidence are the horizon, and \(\tau\) collapses that evidence into ordered fragments when the leader is available.

\paragraph{Denotational reading.}
For Cordial Miners, \(h \sqsubseteq h'\) means that \(h'\) extends the blocklace with additional blocks, acknowledgments, ratification edges, and equivocation evidence while preserving the partial order already denoted by \(h\). The admissibility predicate \(\mathrm{Adm}\) keeps only refinements compatible with the blocklace validity and fault assumptions. The observation \(\mathsf{obs}(h)\) is the ordered fragment emitted by \(\tau\), when a final leader is visible. If no final leader exists in the current wave, the horizon does not yet determine an ordered fragment; ordering is deferred until later blocklace evidence supplies a final-leader witness.

\paragraph{Comparative reading.}
Cordial Miners uses a blocklace, a DAG-like structure, to support dissemination, equivocation exclusion, and ordering \cite{keidar-cordial-miners-2023}. It organizes rounds into waves and uses final leaders ratified by supermajority evidence. In the normal form:
\begin{itemize}[leftmargin=1.2em]
\item \textbf{carrier:} blocklace events, acknowledgments, ratification edges, equivocation evidence, and wave structure;
\item \textbf{legitimacy:} ratification and super-ratification by a supermajority of blocks;
\item \textbf{horizon:} wave length and the presence or absence of a final leader in a wave;
\item \textbf{collapse:} the deterministic function $\tau$ traverses from final leaders and emits newly ordered fragments;
\item \textbf{discarded distinctions:} blocklace concurrency after $\tau$ outputs a total order fragment;
\item \textbf{repair mechanism:} if no final leader exists in a wave, ordering is deferred rather than forced.
\end{itemize}

Cordial Miners preserves history through the wave, collapses at final leaders, and defers when no final leader is visible.

The blocklace perspective is broader than consensus: it treats partially ordered communication evidence as a reusable distributed object, with total-order consensus appearing as one possible collapse policy over that object \cite{almeida-shapiro-blocklace-2024}.

\subsection{Mysticeti-C}

Consider a proposer slot \(s\). The carrier is the uncertified slot DAG: proposals, signed blocks, references, anchors, equivocation evidence, and classifications of later slots. The legitimacy relation extracts support, skip, and implicit certificate patterns for \(s\) from the DAG itself rather than from explicit per-block certificates. The direct horizon is intentionally short, and the direct collapse rule classifies \(s\) as to-commit, to-skip, or undecided.

If the slot remains undecided, the indirect horizon consults later anchor and causal evidence. The protocol does not treat the direct horizon as sufficient. It marks uncertainty at the slot frontier and later reads preserved evidence to repair that uncertainty.

\paragraph{Protocol vocabulary.}
Slot, proposal, signed block, support pattern, skip pattern, implicit certificate pattern, anchor, direct rule, indirect rule, to-commit, to-skip, undecided.

\paragraph{Collapse-policy specification.}
\begin{specdisplay}
\mathcal{H}_{\mathrm{Mysticeti}} &:& \specdesc{uncertified slot DAG, proposals, signed blocks,\\ anchors, equivocation evidence} \\
\mathcal{E}_{\mathrm{Mysticeti}} &:& \specdesc{support, skip, and implicit certificate\\ patterns for a proposer slot} \\
\mathcal{V}_{\mathrm{Mysticeti}} &:& \specdesc{direct horizon or later anchor horizon} \\
\mathcal{P}_{\mathrm{Mysticeti}} &:& \{\mathrm{to\mbox{-}commit},\mathrm{to\mbox{-}skip},\mathrm{undecided}\} \\[0.4em]
L_{\mathrm{Mysticeti},s}(h) &\mathrel{:=}&
\left\langle
\begin{array}{l}
\{v \mid \exists b \in \mathrm{blocks}(h).\; v \text{ authored } b \text{ and } b \text{ supports } s\},\\
\{v \mid \exists b \in \mathrm{blocks}(h).\; v \text{ authored } b \text{ and } b \text{ skips } s\}
\end{array}
\right\rangle \\
R_{\mathrm{Mysticeti},\mathrm{dir}}(e) &\mathrel{:=}&
\text{the part of } e \text{ visible to the direct rule} \\
R_{\mathrm{Mysticeti},\mathrm{ind}}(e) &\mathrel{:=}&
\text{the part of } e \text{ visible through later anchors} \\
C_{\mathrm{Mysticeti}}(u) &\mathrel{:=}&
\begin{cases}
\mathrm{to\mbox{-}commit} & \text{if } u \text{ contains the required implicit certificate pattern for } s,\\
\mathrm{to\mbox{-}skip} & \text{if } u \text{ contains the required skip pattern for } s,\\
\mathrm{undecided} & \text{otherwise.}
\end{cases}
\end{specdisplay}

\paragraph{Protocol reading.}
The slot DAG is the carrier, support/skip and implicit certificate patterns are the legitimacy witnesses, the direct rule is the first horizon, and the indirect anchor rule repairs undecided slots later.

\paragraph{Denotational reading.}
For Mysticeti-C, \(h \sqsubseteq h'\) means that \(h'\) extends the uncertified slot DAG with later signed blocks, anchors, support/skip patterns, implicit certificate patterns, and equivocation evidence while preserving the causal links visible in \(h\). The admissibility predicate \(\mathrm{Adm}\) restricts refinements to DAG extensions that satisfy block validity, pattern validity, quorum thresholds, and the protocol's equivocation bounds. The observation \(\mathsf{obs}_s(h)\) for a slot \(s\) is its classification: to-commit, to-skip, or still undefined for prefix advancement. The direct rule asks whether the short horizon already determines \(\mathsf{obs}_s\). The \emph{undecided} outcome means exactly that admissible refinements can still change the slot observation. The indirect rule widens the horizon by reading later anchor evidence until the observation becomes invariant, or until the prefix frontier remains blocked.

\paragraph{Comparative reading.}
Mysticeti-C names both the short-horizon rule and the later recovery path. It operates on a DAG of round-indexed proposer slots and classifies each slot as \emph{to-commit}, \emph{to-skip}, or \emph{undecided} \cite{babel-mysticeti-2025}.

In the normal form:
\begin{itemize}[leftmargin=1.2em]
\item \textbf{carrier:} round-indexed DAG slots, signed blocks, causal links, anchors, and observed slot classifications;
\item \textbf{legitimacy:} support, skip, and implicit certificate patterns read from later DAG blocks;
\item \textbf{horizon:} the direct-rule visibility surface for a slot, extended later by anchors;
\item \textbf{collapse:} a slot is classified as to-commit or to-skip, and the ordered output advances until the first undecided slot;
\item \textbf{discarded distinctions:} causal evidence outside the direct rule's immediate horizon is not part of the direct classification surface;
\item \textbf{repair mechanism:} the indirect rule consults later anchor evidence, causal links, and implicit certificate patterns; undecided slots prevent unsafe prefix collapse.
\end{itemize}

The direct rule intentionally uses a short evidence horizon. If that horizon is not sufficient, the protocol does not recover information from the emitted prefix. It either refuses to collapse by leaving the slot undecided, or it uses later preserved DAG evidence through an indirect anchor rule. The indirect rule pays collapse debt created by the limited direct horizon.

Mysticeti does not merely add a protocol fallback. Progress cannot always be computed from the information visible to the direct rule, so the later rule reads more of the communication carrier.

Mysticeti also pipelines the collapse frontier: slot decisions are evaluated in a steady flow rather than only at isolated waves. This makes Mysticeti closer to HotStuff's chained/pipelined style, but the pipeline runs over a DAG slot structure rather than a single leader chain. Its \emph{to-skip} rule is a negative collapse rule: the protocol can advance by deciding that a slot should not contribute to the prefix, not only by positively committing a slot. The \emph{undecided} state marks the remaining collapse debt.

\subsection{Worked derivation: one Mysticeti-C slot}

This subsection instantiates the preceding collapse-policy specification for one slot in Mysticeti-C. It is not an additional protocol definition or proof; it shows how the normal form names the direct rule, the undecided state, and the later indirect rule.

Consider a proposer slot \(s\) in round \(r\). Let \(H_t\) be the local DAG visible to a validator at logical time \(t\). The relevant events include proposals for \(s\), signed blocks in later rounds, references between blocks, support or non-support for proposals, implicit certificate patterns, observed equivocations, and the classifications already assigned to later slots.

\[
H_t
\xrightarrow{\;L_s\;}
E_t(s)
\xrightarrow{\;R_{\mathrm{dir}}\;}
E_t(s)|_{R_{\mathrm{dir}}}
\xrightarrow{\;C_{\mathrm{dir}}\;}
\{\mathrm{to\mbox{-}commit},\mathrm{to\mbox{-}skip},\mathrm{undecided}\}.
\]

Here \(L_s\) extracts support patterns relevant to slot \(s\), such as enough later blocks witnessing an implicit certificate pattern, or a skip pattern showing that the slot should not contribute to the output. The direct horizon \(R_{\mathrm{dir}}\) is intentionally short: it asks whether the local DAG already contains enough nearby evidence to classify \(s\). The direct collapse rule \(C_{\mathrm{dir}}\) has three outcomes:
\begin{itemize}[leftmargin=1.2em]
\item \textbf{to-commit:} the direct horizon contains sufficient support for a proposal in \(s\);
\item \textbf{to-skip:} the direct horizon contains sufficient evidence that no proposal in \(s\) should contribute to the output;
\item \textbf{undecided:} the direct horizon does not justify either positive or negative collapse.
\end{itemize}

The third outcome is a refusal to project across insufficient evidence. It is not a terminal prefix decision. If slot \(s\) remains undecided, the protocol consults a later horizon:
\[
H_t
\xrightarrow{\;L_s\;}
E_t(s)
\xrightarrow{\;R_{\mathrm{ind}}\;}
E_t(s)|_{R_{\mathrm{ind}}}
\xrightarrow{\;C_{\mathrm{ind}}\;}
\{\mathrm{to\mbox{-}commit},\mathrm{to\mbox{-}skip},\mathrm{undecided}\}.
\]

The indirect horizon \(R_{\mathrm{ind}}\) includes later committed-anchor information, causal links, and implicit certificate patterns. If the relevant later anchor is itself undecided, the earlier slot may remain undecided. If a later committed anchor causally references a certificate pattern over \(s\), the slot can be classified to-commit. A to-skip classification is justified only relative to that committed-anchor horizon: the later anchor must be sufficient for the protocol's indirect rule to conclude that no relevant certificate pattern for \(s\) is present. Thus the indirect rule does not repair the prefix after the fact. It reads evidence that was preserved in the DAG but not visible, or not decisive, inside the direct horizon.

Two local histories can therefore agree on the currently emitted prefix while differing in later DAG evidence about an earlier undecided slot. A pure prefix output cannot distinguish them. Mysticeti-C keeps the communication carrier available long enough for the indirect rule to distinguish them later. This is the concrete form of collapse debt: the short direct horizon buys latency when evidence is clear, and the undecided/indirect path preserves safety when it is not.

\section{Synthesis}

\subsection{Shared semantic shape}

The main case studies expose a repeated shape. Each protocol accumulates an order-2 communication carrier, extracts decision-grade evidence, restricts attention to a horizon, and then either collapses to an order-1 output or applies a repair/deferral rule. The terminal artifact is the endpoint of this semantic operation, not the object from which the safety argument begins. The appendix checks the same shape across a wider protocol landscape. The surface terminology differs; the semantic control problem is shared.

\paragraph{Carrier first.}
Lamport clocks and the Byzantine Generals Problem are the base cases. Lamport starts from causally ordered events and projects them to a total order only when a total order is needed. Byzantine Generals starts from local report histories and shows that collapse to a common value is unsafe until loyal participants have compatible interpretations of who said what. In both cases, the terminal value or order is not the primitive object. The communication carrier is.

\paragraph{Quorum recovery.}
Paxos uses majority evidence to protect a value across later attempts to collapse. It carries prior accepted values through higher ballots, so the repair rule is not to restart history but to recover the evidence that may already constrain future collapse. Raft, treated in the appendix, gives the replicated-log version of the same crash-fault pattern.

\paragraph{Certificate compression.}
HotStuff moves the argument into Byzantine quorum evidence. It compresses multi-phase support into quorum-certificate chains, making view change and pipelining cheaper. PBFT, treated in the appendix, is the phase-certificate predecessor. The shared pattern is not the specific phase structure; it is portable threshold evidence that narrows what later views may safely forget.

\paragraph{Alternative legitimacy carriers.}
The appendix shows that legitimacy need not always be a fixed validator-set quorum certificate. Stellar derives quorum evidence from local quorum slices. Algorand uses VRF-selected committees. Ouroboros uses stake-weighted leader evidence and chain stability. GRANDPA and Casper FFG overlay finality votes on a block tree or checkpoint tree. Avalanche and Snowman accumulate sampled preference evidence until a choice becomes stable enough to expose.

\paragraph{Semantic closure.}
CBC Casper states the collapse condition most directly. A protocol state is a justified message carrier, and a decision is safe only when the estimator is invariant over all admissible future states. In this family, the horizon is not just a round or view. It is an analysis of future extensions of the carrier.

\paragraph{DAG interpretation.}
Cordial Miners and Mysticeti-C make the modern DAG/blocklace version of the carrier explicit in the main text. Cordial Miners walks backward from final leaders through a blocklace. Mysticeti-C classifies proposer slots with direct and indirect horizons. The appendix adds DAG-Rider, Hashgraph, Aleph, and Narwhal/Tusk/Bullshark as broader instances. These protocols differ in commit rules, but they share the same semantic move: preserve a partial order long enough for a later local interpretation to emit a prefix.

\subsection{Cordial Miners and Mysticeti-C}

Cordial Miners and Mysticeti-C are especially close in the collapse view. Both use later partially ordered communication evidence to resolve earlier ordering ambiguity. In Cordial Miners, a later super-ratified leader anchors finality, and the ordering function $\tau$ walks backward through ratified structure. In Mysticeti-C, later anchor evidence, causal links, and implicit certificate patterns can indirectly decide earlier slots when the direct rule is insufficient.

Their difference is the granularity of collapse:
\begin{itemize}[leftmargin=1.2em]
\item \textbf{Cordial Miners:} sparse wave-anchor collapse. If a final leader exists, a fragment is finalized; if no final leader exists, ordering is deferred.
\item \textbf{Mysticeti-C:} pipelined slot-frontier collapse. Each slot may be classified as to-commit, to-skip, or undecided, and the output frontier advances until the first undecided slot.
\end{itemize}

Thus Mysticeti-C can be read as a finer-grained slot-level analogue of the same backward-evidence pattern, with two extra collapse controls. Multiple proposer/leader slots per round densify the evidence horizon, and the skip rule supplies a negative decision that Cordial Miners does not expose at the same per-slot level. Pipeline overlap then turns these slot decisions into a continuous frontier, analogous in spirit to HotStuff's pipelined QC chain but lifted into a DAG carrier.

The apparent terminology gap is smaller than it looks. ``Super-ratified leader,'' ``anchor,'' ``direct rule,'' ``indirect rule,'' and ``skip pattern'' are different protocol terms for related choices about when later order-2 evidence is strong enough to settle earlier history, and whether failure to settle should produce deferral, skipping, or an undecided frontier.

\subsection{Axes of difference}

\paragraph{Where legitimacy lives.}
Legitimacy may be implicit in causality, as in Lamport ordering; reconstructed through relay or signatures, as in Byzantine Generals; supplied by majority intersection, as in Paxos and Raft; compressed into certificates, as in PBFT and HotStuff; derived from slices or committee credentials, as in Stellar and Algorand; accumulated as confidence, as in Snowman; checked as estimator safety, as in CBC Casper; or read from DAG/blocklace structure, as in the DAG protocols.

\paragraph{How wide the horizon is.}
Leader protocols tend to choose narrow horizons: a ballot, a term, a view, or a QC chain. DAG protocols often retain wider local visibility: waves, witness rounds, certified DAG windows, anchors, or blocklace waves. Wider horizons preserve more order-2 distinctions but usually increase dissemination, storage, or interpretation cost.

\paragraph{When collapse is refused.}
The repair path is the clearest diagnostic of the chosen horizon. Paxos starts a higher ballot. Raft elects a new leader and repairs log suffixes. PBFT changes view. HotStuff rotates leaders and carries the highest QC. CBC Casper waits for more justified messages or fault evidence. DAG protocols keep extending the DAG, choose later waves, decide witness fame later, defer a final leader, or leave a Mysticeti slot undecided.

\paragraph{What gets discarded.}
Every collapse hides distinctions. Logical clocks hide concurrency by imposing a total order. Paxos hides losing proposals. Raft hides failed leader histories and overwritten uncommitted suffixes. PBFT and HotStuff hide uncertified branches and timing histories. DAG protocols hide event/block concurrency once they expose a total order fragment. The question is not whether information is discarded, but whether the discarded distinctions can still affect the correctness property being claimed.

\subsection{Communication Surface and Compression}

The same collapse view also separates two quantities that are often conflated: how much order-2 evidence exists, and how expensively that evidence is disseminated.

\paragraph{Full-gossip DAG surface.}
In the usual DAG style, each validator produces blocks or vertices and disseminates them broadly so that other validators can independently observe the causal structure. If $n$ validators each broadcast to $n$ validators in a round, the message surface is quadratic, $O(n^2)$, before counting payload size, signatures, or retransmission. The benefit is rich local visibility: each validator can interpret much of the DAG without relying on a single relay.

\paragraph{Metric caveat.}
The asymptotic vocabulary in the literature is easy to misread because papers count different units. A DAG round may have an $O(n^2)$ message surface because $O(n)$ validators each send to $O(n)$ recipients, yet the amortized cost per block can be reported as $O(n)$ because the round carries $O(n)$ blocks. Hashgraph shifts much of the voting cost into local computation by gossiping the causal history from which virtual votes are computed \cite{baird-hashgraph-2016}. Reliable Broadcast can add another layer: Cordial Miners notes that Bracha Reliable Broadcast has $O(n^2)$ message complexity per broadcast message, and that DAG-Rider and Bullshark use Reliable Broadcast as a building block \cite{keidar-cordial-miners-2023}. Mysticeti-C states the tradeoff even more directly: it prioritizes low latency and DAG amortization while embracing a higher, cubic cost compared with linear-communication protocols \cite{babel-mysticeti-2025}. Thus the precise claim is not that every DAG protocol has the same complexity, but that full local DAG visibility is bought by a broader communication surface unless the protocol compresses or relays evidence.

\paragraph{Leader/QC compression surface.}
HotStuff-style protocols compress evidence through a leader path \cite{yin-hotstuff-2019}. Validators send votes to a leader or proposer, the leader aggregates them into a quorum certificate, and the certificate is broadcast back. With threshold signatures or compact aggregation, the message surface per step is linear, $O(n)$: $O(n)$ votes to the leader and $O(n)$ dissemination from the leader. The semantic price is that visibility is leader-mediated. Evidence not carried by the leader path is not part of the immediate collapse horizon.

\paragraph{DAGs can use the same compression.}
A DAG-shaped protocol is not forced to use full-gossip dissemination. It can choose a dissemination leader or relay for a round, slot, or wave. Validators send DAG blocks, references, votes, or availability evidence to that leader; the leader recomposes, certifies, or aggregates the evidence; and then broadcasts the compressed result to the rest of the validators. The resulting message count can be linear in $n$, much like HotStuff, while retaining a DAG-shaped object at the protocol boundary.

But this is no longer the same semantic point as full-gossip DAG exposure. It is a leader-compressed DAG:
\begin{itemize}[leftmargin=1.2em]
\item \textbf{message count:} can move from $O(n^2)$ broad dissemination toward $O(n)$ leader relay;
\item \textbf{evidence size:} can be compressed by quorum certificates, threshold signatures, bitmaps, or aggregate availability witnesses;
\item \textbf{visibility:} becomes filtered through the leader or relay path;
\item \textbf{risk:} faulty, slow, or censoring leaders can hide or delay parts of the DAG unless fallback or view-change rules recover them;
\item \textbf{semantic effect:} the protocol preserves DAG syntax but moves its collapse horizon closer to a leader-mediated BFT design.
\end{itemize}

Thus the main distinction is not ``DAG versus linear BFT.'' It is the degree and location of evidence compression before collapse. Full-gossip DAG protocols and leader/QC protocols sit near opposite ends of this spectrum, while many practical designs occupy intermediate points: they retain a DAG-shaped carrier, but vary how dissemination, visibility, anchors, certificates, relays, reputation, or leaders compress the evidence used for collapse.

\subsection{Output-First and Communication-First Readings}

The formal statements of the protocols are not at issue. The difference is the direction from which they are read.

\paragraph{Output-first reading.}
From the output-first direction, each protocol appears to solve a different problem with different terminology: Paxos ballots, Raft terms, PBFT phases, Stellar quorum slices, HotStuff chains, Tendermint/Streamlet rounds, Algorand sortition committees, Ouroboros stable prefixes, GRANDPA and Casper FFG finality justifications, Avalanche/Snowman confidence thresholds, Jolteon/Ditto fallback, CBC Casper safety oracles, DAG waves, Hashgraph virtual voting, Aleph poset ordering, Cordial blocklace traversal, and Mysticeti direct and indirect rules. The common object is hidden because the analysis starts at the value or prefix.

\paragraph{Communication-first reading.}
From the communication-first direction, the protocols discussed here are variants of one normal form. They all ask:
\begin{quote}
Given a growing communication carrier, when is enough evidence visible to safely stop preserving alternatives and emit an order-1 output?
\end{quote}

This question explains why fallback rules keep appearing. A fallback is not an accidental complication. It is evidence that the original horizon sometimes cannot support the requested collapse.

\paragraph{Mysticeti as the diagnostic case.}
Mysticeti's direct and indirect rules make the phenomenon explicit. The direct rule uses one visibility horizon. The indirect rule uses later anchor evidence. The undecided state marks the refusal to project a prefix across insufficient evidence. Thus the protocol itself reveals the order distinction: progress may require information not present in the direct rule's chosen carrier.

\section{Related Work and Positioning}

\paragraph{Impossibility and progress assumptions.}
FLP gives the classical asynchronous crash-failure boundary for deterministic consensus \cite{fischer-lynch-paterson-1985}. Dwork, Lynch, and Stockmeyer show how partial synchrony changes the progress setting \cite{dwork-lynch-stockmeyer-1988}. Chandra and Toueg introduce failure detectors as additional information that can make consensus solvable in asynchronous crash-failure systems \cite{chandra-toueg-1996}. Ben-Or, Rabin, Canetti--Rabin, and HoneyBadgerBFT show the randomized asynchronous lineage: progress is obtained not by timing assumptions, but by random choices, common coins, reliable broadcast, and asynchronous agreement structure \cite{benor-free-choice-1983,rabin-randomized-generals-1983,canetti-rabin-1993,miller-honeybadger-2016}. In this vocabulary, these works characterize when the terminal collapse rule can be required, expected, or made to terminate with probability one.

\paragraph{Set agreement and topology.}
Chaudhuri's set agreement exposes output width as a parameter: consensus is the \(k=1\) case, while \(k\)-set agreement permits bounded residual disagreement \cite{chaudhuri-set-consensus-1990}. Borowsky--Gafni resilient simulation generalizes the obstruction through crash resilience \cite{borowsky-gafni-1993}. Herlihy--Shavit and Saks--Zaharoglou give the topological form: protocol complexes record views/executions, task complexes record outputs, and solvability depends on an appropriate map between them \cite{herlihy-shavit-topology,saks-zaharoglou-1993}. These works are the formal background for the carrier-to-output reading.

\paragraph{Classical quorum/certificate protocols.}
Paxos, Raft, PBFT, HotStuff, Tendermint, Streamlet, Jolteon, and Ditto are not treated here as equivalent protocols. They differ in fault model, leader structure, view change, responsiveness, fallback path, and proof obligations \cite{lamport-paxos-1998,lamport-paxos-simple-2001,ongaro-raft-2014,castro-liskov-1999,yin-hotstuff-2019,kwon-tendermint-2014,chan-shi-streamlet-2020,gelashvili-jolteon-ditto-2022}. Their shared legitimacy backbone is quorum or certificate evidence: majority evidence in the crash-fault case, supermajority evidence in the Byzantine case, and corresponding threshold witnesses in optimized variants. The protocol-specific variation is how that evidence is carried, bounded by horizons, compressed, adapted, and projected into a terminal value, log position, or prefix.

\paragraph{Federated and committee-selected quorums.}
Stellar and Algorand change how the quorum carrier is chosen \cite{mazieres-stellar-2015,gilad-algorand-2017}. Stellar lets each participant choose quorum slices, so legitimacy is induced by overlapping local trust choices. Algorand uses VRF sortition to select committees privately for each agreement step, so eligibility evidence travels with the vote. Both still collapse communication evidence to a value or block, but the legitimacy layer is no longer simply a fixed global validator quorum.

\paragraph{Proof-of-stake chains and finality gadgets.}
Ouroboros, GRANDPA, and Casper FFG expose another split in the collapse-policy space \cite{kiayias-ouroboros-2017,david-ouroboros-praos-2018,stewart-grandpa-2020,buterin-casper-ffg-2017}. Ouroboros uses proof-of-stake leader election and chain selection to make a stable prefix meaningful under probabilistic/common-prefix security. GRANDPA takes a block tree from a separate production mechanism and applies weighted prevote/precommit evidence to finalize a prefix. Casper FFG uses source-target checkpoint votes and slashing conditions to justify and finalize checkpoints. In each case, the output is a prefix, but the evidence horizon differs: chain growth and stability in Ouroboros, round-local ancestry voting in GRANDPA, and checkpoint-link voting in Casper FFG.

\paragraph{Metastable sampling.}
Avalanche and Snowman use repeated randomized sampling rather than quorum certificates, finality checkpoints, or DAG waves \cite{rocket-avalanche-2019}. The carrier is a history of sampled preferences and confidence counters. Collapse happens when local confidence has accumulated enough support for a choice or chain prefix. The horizon is a confidence threshold over observations, not a fixed round, wave, or certificate depth.

\paragraph{Estimator-based blockchain consensus.}
CBC Casper starts from protocol states, estimators, admissible futures, and safety of estimates \cite{zamfir-casper-tfg-2017}. Its blockchain instance uses GHOST as an estimator, but its finality condition is not simply ``follow GHOST.'' A block is finalized only when it is safe: compatible with the estimator over all admissible future protocol states under the fault threshold. The result is semantic closure rather than only a mechanism for reaching a prefix.

\paragraph{DAG-based consensus and atomic broadcast.}
DAG-Rider, Hashgraph, Aleph, Narwhal/Tusk, Bullshark, Cordial Miners, and Mysticeti-C make communication history more explicit than single-chain protocols \cite{keidar-dagrider-2021,baird-hashgraph-2016,gagol-aleph-2019,danezis-narwhal-tusk-2022,spiegelman-bullshark-2022,keidar-cordial-miners-2023,babel-mysticeti-2025}. DAG-Rider separates reliable-broadcast DAG construction from local ordering. Hashgraph records gossip-about-gossip and computes votes virtually from the resulting event DAG. Aleph builds a partially ordered set and computes a common total-order extension with randomized asynchronous machinery. Narwhal separates dissemination/storage of causal transaction histories from ordering. Cordial Miners uses a blocklace for dissemination, equivocation exclusion, and ordering. Mysticeti-C exposes direct and indirect classification of slots. These protocols are the closest operational relatives of the collapse-policy view.

\paragraph{Position of this paper.}
This paper is not a replacement for protocol-specific correctness proofs or performance analyses. It is a semantic comparison layer over them. Its normal form asks which evidence is retained, which horizon is inspected, which projection is emitted, and which repair path exists when the horizon is insufficient.

\section{Discussion}

The collapse view does not imply that all protocols are equally good. It gives a better basis for comparing them.

\begin{itemize}[leftmargin=1.2em]
\item A protocol may be faster because it collapses earlier.
\item A protocol may be safer under equivocation because it preserves more causal evidence.
\item A protocol may reduce communication by compressing evidence through leaders or quorum certificates.
\item A protocol may improve liveness by adding randomness, timing assumptions, or deferral rules.
\item A protocol may look DAG-like syntactically while semantically behaving like leader-mediated evidence compression.
\end{itemize}

These are different achievements. Calling all of them ``consensus improvements'' hides the tradeoff. The normal form separates them: legitimacy, horizon, compression, collapse, and repair.

\paragraph{Leader compression.}
Leader-based protocols often reduce communication surfaces by concentrating evidence flow. That is not merely an engineering optimization. It changes which participants see which order-2 distinctions before collapse. A leader can coordinate a prefix cheaply, but the price is stronger dependence on leader-mediated visibility.

\paragraph{DAG exposure.}
DAG-based protocols preserve more communication history explicitly. That can make local interpretation cheaper once evidence is visible, and can support later fallback rules that read causal structure not available to shorter horizons. The price is usually larger dissemination, storage, or verification surface.

\paragraph{Design implication.}
Protocol specifications and comparisons benefit from stating the collapse policy explicitly: what evidence makes communication decision-grade, what history is retained, what horizon is inspected, what distinctions are discarded at output, and what repair rule applies when the horizon is insufficient.

\section{Future Work}

The collapse-policy view also suggests a design space for new protocols. Instead of beginning with a fixed message pattern or commit rule, a protocol can begin by choosing its communication carrier \(H\), then choosing legitimacy \(L\), horizon \(R\), collapse \(C\), and repair/deferral rules.

One direction is adaptive horizons: collapse early when the carrier contains clean decision-grade evidence, widen the horizon when ambiguity or equivocation remains, and leave explicit collapse debt when neither positive nor negative collapse is justified. Another is carrier-preserving finality, where the protocol finalizes a causally closed subhistory and exposes a linear prefix only as a projection. A third is leader-compressed DAG design: retain a DAG-shaped semantic carrier while using quorum certificates, threshold aggregation, or relays to reduce the communication surface.

These directions treat protocol design as semantic resource management. The resources are not only messages, signatures, and latency, but also preserved distinctions, visible evidence, and valid points of collapse.

\section{Conclusion}

Consensus protocols are usually described by their final artifact. This paper describes them by the semantic operation that produces that artifact: lawful collapse of a communication carrier into a value, log, or prefix.

The resulting picture is more uniform than the protocol vocabulary suggests. Lamport clocks, Byzantine oral/signed reports, Paxos ballots, Raft terms, PBFT phases, Stellar quorum slices, HotStuff quorum-certificate chains, Tendermint/Streamlet rounds, Algorand committee certificates, Ouroboros stable-prefix rules, GRANDPA and Casper FFG finality justifications, Avalanche/Snowman confidence thresholds, Jolteon/Ditto fallback, CBC Casper safety oracles, DAG-Rider waves, Hashgraph virtual voting, Aleph poset ordering, Cordial Miners ratification, and Mysticeti direct/indirect rules are read here as managing related versions of the same underlying problem: when does a growing order-2 communication carrier contain enough decision-grade evidence for safe projection to an order-1 output?

This does not refute FLP and does not introduce a new impossibility theorem. It relocates the insight. FLP constrains deterministic guaranteed terminal collapse under full asynchrony and one crash failure. Chaudhuri, Borowsky--Gafni, and topological distributed computing show that the obstruction scales: the issue is not merely one binary decision, but the relationship between adversarially preserved alternatives and the width/topology of the output carrier. The collapse view explains why practical protocols keep adding horizons, certificates, leaders, randomness, timeouts, anchors, waves, and fallback rules: they are mechanisms for deciding when history is allowed to stop mattering.

The practical consequence is a specification method. Instead of asking only which protocol agrees fastest, we should ask what evidence it accepts, what history it preserves, what it compresses, when it collapses, and how it repairs insufficient evidence. Nontrivial distributed consensus protocols decide when a growing communication carrier contains enough evidence to project safely to a value, log, or prefix.

\section*{Acknowledgments}

I thank Ehud Shapiro for correspondence regarding Cordial Miners and its relation to Mysticeti-style DAG consensus, and Kushal Babel for clarifying the role of the direct and indirect rules in Mysticeti. Any remaining errors or interpretations are my own.


\appendix
\section{Broader Protocol Instantiations}

The main text develops the semantic argument through cases that expose distinct roles: causal projection, carrier enrichment, quorum recovery, certificate compression, admissible-future safety, and DAG collapse debt. The following entries are not secondary in protocol importance. They are secondary in the proof strategy: they instantiate the same normal form across a broader protocol landscape once the denotational reading has been established.

The appendix entries are schematic normal-form readings, not standalone correctness proofs of the protocols.

\subsection{HoneyBadgerBFT and randomized asynchronous agreement}

Randomized asynchronous agreement changes the progress side of the collapse policy without adding timing assumptions. Ben-Or's protocol uses private random choices to prevent the adversary from deterministically preserving bivalence forever \cite{benor-free-choice-1983}. Rabin's Byzantine agreement and Canetti--Rabin's optimally resilient asynchronous Byzantine agreement strengthen this line with shared/random-coin machinery and cryptographic structure \cite{rabin-randomized-generals-1983,canetti-rabin-1993}. HoneyBadgerBFT provides the concrete ACS/batch carrier: encrypted inputs are reliably broadcast, binary agreement and a common coin choose the batch, and threshold decryption opens only the selected outputs \cite{miller-honeybadger-2016}.

\paragraph{Protocol vocabulary.}
Epoch, encrypted input, reliable broadcast, binary agreement, common coin, ACS instance, decryption share, batch.

\paragraph{Collapse-policy specification.}
\begin{specdisplay}
\mathcal{H}_{\mathrm{HB}} &:& \specdesc{enc. inputs, RBC transcripts, BA messages, coins,\\ decryption shares} \\
\mathcal{E}_{\mathrm{HB}} &:& \specdesc{RBC outputs, BA decisions,\\ coin/decryption evidence} \\
\mathcal{V}_{\mathrm{HB}} &:& \specdesc{one epoch's ACS/decryption horizon} \\
\mathcal{P}_{\mathrm{HB}} &:& \text{ordered transaction batch} \\[0.4em]
L_{\mathrm{HB}}(h) &\mathrel{:=}&
\{(i,x) \mid \mathrm{RBC}_i \text{ delivers } x \text{ in } h\} \\
R_{\mathrm{HB},e}(a) &\mathrel{:=}&
\{(i,x) \in a \mid \mathrm{BA}_e(i)=1\} \\
C_{\mathrm{HB}}(u) &\mathrel{:=}&
\begin{cases}
\mathrm{order}(\mathrm{decrypt}(u)) & \text{if decryption shares are valid},\\
\text{undefined} & \text{otherwise.}
\end{cases}
\end{specdisplay}

\paragraph{Protocol reading.}
Reliable broadcast makes encrypted inputs available, binary agreement/ACS chooses which inputs enter the epoch batch, and threshold decryption opens only the selected batch. The common coin is not a terminal output; it is repair evidence for making the ACS horizon progress in asynchrony.

\paragraph{Comparative reading.}
In the normal form:
\begin{itemize}[leftmargin=1.2em]
\item \textbf{carrier:} encrypted inputs, reliable-broadcast delivery evidence, binary-agreement messages, coin shares, and decryption shares;
\item \textbf{legitimacy:} reliable-broadcast delivery, ACS agreement evidence, and valid threshold decryption shares;
\item \textbf{horizon:} one epoch's ACS instance and decryption step;
\item \textbf{collapse:} decrypt and order the ACS-selected inputs as a batch;
\item \textbf{discarded distinctions:} inputs not selected for the epoch and schedules that did not affect the ACS output;
\item \textbf{repair mechanism:} common-coin and repeated asynchronous agreement steps until ACS completes.
\end{itemize}

The broader Ben-Or/Rabin/Canetti--Rabin lineage explains why this progress path works: randomness prevents the adversary from deterministically preserving the same ambiguity forever. The HoneyBadger-style spec above keeps the carrier concrete.

\subsection{Raft}

Raft makes the leader path explicit. A leader in term \(T\) wants to commit log index \(j\). The carrier contains RequestVote evidence, AppendEntries messages, follower logs, match indices, current terms, and commit indices. The legitimacy condition is majority election plus majority replication. Election restriction and log matching ensure that later leaders cannot safely overwrite committed entries.

The horizon is the leader's current term, its local log, and the follower match-index frontier it learns from AppendEntries responses. When a log entry is replicated on a majority under Raft's rules, the leader may advance the commit index. If a leader fails, a new election reestablishes the evidence horizon; if a follower has a conflicting suffix, AppendEntries conflict repair overwrites uncommitted suffixes while preserving the committed prefix. That is the collapse policy in concrete form.

\paragraph{Protocol vocabulary.}
Term, RequestVote, leader election, AppendEntries, follower log, match index, commit index.

\paragraph{Collapse-policy specification.}
\begin{specdisplay}
\mathcal{H}_{\mathrm{Raft}} &:& \specdesc{terms, votes, AppendEntries, logs, match indices,\\ commit indices} \\
\mathcal{E}_{\mathrm{Raft}} &:& \specdesc{majority election and replication evidence} \\
\mathcal{V}_{\mathrm{Raft}} &:& \specdesc{current-term leader horizon and\\ match-index frontier} \\
\mathcal{P}_{\mathrm{Raft}} &:& \text{committed prefix} \\[0.4em]
L_{\mathrm{Raft}}(h) &\mathrel{:=}&
\{(i,j,T) \mid \text{follower } i \text{ stores log index } j \text{ in term } T \text{ in } h\} \\
R_{\mathrm{Raft},T}(e) &\mathrel{:=}&
\{(i,j,T') \in e \mid T'=T \text{ and } i \text{ is visible to the leader}\} \\
C_{\mathrm{Raft}}(v) &\mathrel{:=}&
\max\{j \mid |\{i \mid (i,j,T) \in v\}| \text{ is a majority}\}.
\end{specdisplay}

\paragraph{Protocol reading.}
The term and match-index frontier are the visible cut, the commit index is the collapse point, and conflict repair only touches uncommitted suffixes.

\paragraph{Comparative reading.}
Raft reorganizes the Paxos lineage around understandability by separating leader election, log replication, and safety \cite{ongaro-raft-2014}. Its terminal output is a replicated log, and its communication carrier is deliberately structured around terms and a leader-maintained prefix.

In the normal form:
\begin{itemize}[leftmargin=1.2em]
\item \textbf{carrier:} terms, votes, AppendEntries messages, replicated log entries, commit indices, and leader-election evidence;
\item \textbf{legitimacy:} majority election and majority replication;
\item \textbf{horizon:} current term, leader log, and follower match indices;
\item \textbf{collapse:} a leader commits a log entry once it is replicated on a majority under Raft's safety rules;
\item \textbf{discarded distinctions:} concurrent client requests, failed leader proposals, and overwritten uncommitted suffixes;
\item \textbf{repair mechanism:} election restriction, log matching, leader completeness, and conflict repair in AppendEntries.
\end{itemize}

Raft makes leader-mediated collapse explicit. It narrows the communication carrier so that one leader orders commands into a log, while majority evidence protects already-committed prefixes across leader changes.

\subsection{PBFT}

PBFT fixes a request \(m\) and sequence number \(n\) inside a view \(v\). The carrier contains the primary's pre-prepare, prepare votes, commit votes, checkpoints, and any view-change evidence. The legitimacy relation is Byzantine quorum evidence over the same request and sequence number. The horizon is the current view's pre-prepare/prepare/commit phase sequence.

Once enough prepare and commit evidence is visible, the protocol collapses the phase history into a committed log position. If the primary is faulty or the view stalls, replicas do not throw away the safety evidence. They move it into the next view through view change, carrying prepared certificates and checkpoints so that a conflicting request cannot be safely committed later. The repair path is therefore explicit state transfer across views.

\paragraph{Protocol vocabulary.}
Primary, view, pre-prepare, prepare, commit, checkpoint, view change, prepared certificate.

\paragraph{Collapse-policy specification.}
\begin{specdisplay}
\mathcal{H}_{\mathrm{PBFT}} &:& \specdesc{pre-prepare, prepare, commit, checkpoint,\\ and view-change messages} \\
\mathcal{E}_{\mathrm{PBFT}} &:& \specdesc{Byzantine quorum evidence\\ for request/sequence pairs} \\
\mathcal{V}_{\mathrm{PBFT}} &:& \specdesc{current-view phase horizon} \\
\mathcal{P}_{\mathrm{PBFT}} &:& \text{committed log position} \\[0.4em]
L_{\mathrm{PBFT}}(h) &\mathrel{:=}&
\{(r,n,v,\phi) \mid \phi \in \{\mathrm{preprepare},\mathrm{prepare},\mathrm{commit}\} \text{ appears in } h\} \\
R_{\mathrm{PBFT},v}(e) &\mathrel{:=}&
\{(r,n,v',\phi) \in e \mid v'=v\} \\
C_{\mathrm{PBFT}}(u) &\mathrel{:=}&
\{(r,n) \mid u \text{ contains sufficient prepare and commit evidence for } (r,n)\}.
\end{specdisplay}

\paragraph{Protocol reading.}
The phase history is the carrier, quorum evidence is the legitimacy layer, the current view is the horizon, and view change preserves prepared evidence across views.

\paragraph{Comparative reading.}
PBFT-style protocols use a primary/view structure and phase certificates to commit sequence numbers \cite{castro-liskov-1999}. In the normal form:
\begin{itemize}[leftmargin=1.2em]
\item \textbf{carrier:} requests, pre-prepare messages, prepare votes, commit votes, checkpoints, view-change evidence;
\item \textbf{legitimacy:} quorum evidence over a request and sequence number;
\item \textbf{horizon:} the current view and its pre-prepare/prepare/commit phases;
\item \textbf{collapse:} a commit certificate projects the phase history into a log position;
\item \textbf{discarded distinctions:} competing proposals, alternate request orderings, and timing histories that did not enter the committed certificate;
\item \textbf{repair mechanism:} view change carries enough evidence to preserve safety when the current primary or horizon fails.
\end{itemize}

PBFT therefore does not merely ``agree on a value.'' It constructs a bounded evidence horizon and collapses that horizon into an order-1 log position once the phase evidence is sufficient.

\subsection{Stellar}

The Stellar Consensus Protocol changes where quorum legitimacy lives \cite{mazieres-stellar-2015}. Classical BFT protocols usually assume a globally known validator set and fixed quorum threshold. Stellar uses federated Byzantine agreement: each node chooses quorum slices, and system-level quorums arise from overlapping local trust choices. The carrier therefore includes not only votes and nominations, but also the quorum-slice structure needed to decide whether those statements count as quorum evidence at all.

\paragraph{Protocol vocabulary.}
Federated Byzantine agreement, quorum slice, quorum, nomination, ballot, prepared, committed, externalized value.

\paragraph{Collapse-policy specification.}
\begin{specdisplay}
\mathcal{H}_{\mathrm{SCP}} &:& \specdesc{statements, nominations, ballots, quorum slices,\\ local node views} \\
\mathcal{E}_{\mathrm{SCP}} &:& \specdesc{federated quorum evidence and blocking-set\\ evidence for statements} \\
\mathcal{V}_{\mathrm{SCP}} &:& \specdesc{local ballot horizon with visible\\ quorum-slice structure} \\
\mathcal{P}_{\mathrm{SCP}} &:& \text{externalized value for a slot} \\[0.4em]
L_{\mathrm{SCP}}(h) &\mathrel{:=}&
\{(q,x) \mid q \text{ is a quorum under the slice relation in } h \text{ and supports statement } x\} \\
R_{\mathrm{SCP},b}(e) &\mathrel{:=}&
\{(q,x) \in e \mid x \text{ is visible in ballot horizon } b\} \\
C_{\mathrm{SCP}}(u) &\mathrel{:=}&
\begin{cases}
x & \text{if } u \text{ justifies externalizing value } x,\\
\text{undefined} & \text{otherwise.}
\end{cases}
\end{specdisplay}

\paragraph{Protocol reading.}
Quorum slices are part of the communication carrier because they determine which sets of statements count as legitimate evidence. Nomination proposes candidate values, the ballot protocol refines them, and externalization is collapse to the value supported by sufficient federated quorum evidence.

\paragraph{Comparative reading.}
In the normal form:
\begin{itemize}[leftmargin=1.2em]
\item \textbf{carrier:} node statements, quorum slices, nominations, ballots, prepared/committed evidence, and local views;
\item \textbf{legitimacy:} quorum and blocking-set evidence induced by participant-chosen quorum slices;
\item \textbf{horizon:} the local ballot state and visible slice-derived quorum structure;
\item \textbf{collapse:} externalize a value when federated quorum evidence makes it safe;
\item \textbf{discarded distinctions:} unsuccessful nominations, losing ballots, and local trust alternatives not reflected in the externalized value;
\item \textbf{repair mechanism:} continue nomination/ballot progression until federated quorum evidence closes over a value.
\end{itemize}

Stellar makes the legitimacy relation visible inside the carrier. Collapse still emits an order-1 value, but a reader cannot judge the supporting votes without also reading the quorum-slice network that turns local trust choices into federated quorum evidence.

\subsection{Tendermint}

Tendermint sits close to PBFT and HotStuff in the collapse-policy map. Its height/round structure gives a compact blockchain instance of the same design: within one height, each round has a proposer, prevote/precommit evidence, and lock state that determines whether a block can become part of the committed chain \cite{kwon-tendermint-2014}.

\paragraph{Protocol vocabulary.}
Height, round, proposer, prevote, precommit, lock, valid value, committed block.

\paragraph{Collapse-policy specification.}
\begin{specdisplay}
\mathcal{H}_{\mathrm{Tendermint}} &:& \specdesc{proposals, prevotes, precommits, locks, valid values,\\ height/round state} \\
\mathcal{E}_{\mathrm{Tendermint}} &:& \specdesc{quorum prevote/precommit evidence\\ for a block at a height} \\
\mathcal{V}_{\mathrm{Tendermint}} &:& \specdesc{current height and round-local vote horizon} \\
\mathcal{P}_{\mathrm{Tendermint}} &:& \text{committed block for a height} \\[0.4em]
L_{\mathrm{Tendermint}}(h) &\mathrel{:=}&
\{(b,r,\phi) \mid \phi \in \{\mathrm{prevote},\mathrm{precommit}\} \text{ has quorum support for } b \text{ in round } r\} \\
R_{\mathrm{Tendermint},r}(e) &\mathrel{:=}&
\{(b,r',\phi) \in e \mid r'=r \text{ and } \phi \text{ is visible at the current height}\} \\
C_{\mathrm{Tendermint}}(u) &\mathrel{:=}&
\begin{cases}
b & \text{if } u \text{ contains quorum precommit evidence for } b,\\
\text{undefined} & \text{otherwise.}
\end{cases}
\end{specdisplay}

\paragraph{Protocol reading.}
The height and round provide the visibility horizon, prevotes and precommits provide legitimacy, and locks preserve earlier evidence across later proposer attempts.

\paragraph{Comparative reading.}
In the normal form:
\begin{itemize}[leftmargin=1.2em]
\item \textbf{carrier:} proposed blocks, prevotes, precommits, locks, valid values, and height/round state;
\item \textbf{legitimacy:} Byzantine quorum prevote/precommit evidence;
\item \textbf{horizon:} one height and round;
\item \textbf{collapse:} commit a block for the height when quorum precommit evidence is visible;
\item \textbf{discarded distinctions:} losing proposals, skipped leaders, and non-finalized forks;
\item \textbf{repair mechanism:} move to later rounds while preserving lock and valid-value evidence.
\end{itemize}

Tendermint is a clean round-local horizon case. The relevant cut is not just ``a round'' in the abstract; it is the current height and round together with the prevote/precommit evidence and lock state carried from earlier rounds.

\subsection{Streamlet}

Streamlet gives a streamlined pedagogical form of notarized-chain finality \cite{chan-shi-streamlet-2020}. The carrier is a block tree with votes and notarizations. Collapse occurs when a short chain of notarized blocks satisfies the finalization rule.

\paragraph{Protocol vocabulary.}
Epoch, leader, vote, notarized block, notarized chain, finalized block.

\paragraph{Collapse-policy specification.}
\begin{specdisplay}
\mathcal{H}_{\mathrm{Streamlet}} &:& \specdesc{block tree, epoch leaders, votes, notarizations,\\ and ancestry} \\
\mathcal{E}_{\mathrm{Streamlet}} &:& \specdesc{notarized block evidence} \\
\mathcal{V}_{\mathrm{Streamlet}} &:& \specdesc{notarized-chain horizon} \\
\mathcal{P}_{\mathrm{Streamlet}} &:& \text{finalized block prefix} \\[0.4em]
L_{\mathrm{Streamlet}}(h) &\mathrel{:=}&
\{(b,e) \mid b \text{ has quorum votes and is notarized in epoch } e\} \\
R_{\mathrm{Streamlet},e}(a) &\mathrel{:=}&
\{(b,e') \in a \mid e' \leq e \text{ and } b \text{ lies on a visible notarized chain}\} \\
C_{\mathrm{Streamlet}}(u) &\mathrel{:=}&
\{b \mid u \text{ contains a notarized-chain pattern finalizing } b\}.
\end{specdisplay}

\paragraph{Protocol reading.}
Votes produce notarizations, notarizations form the visible chain horizon, and the finalization rule collapses a notarized-chain pattern into a finalized prefix.

\paragraph{Comparative reading.}
In the normal form:
\begin{itemize}[leftmargin=1.2em]
\item \textbf{carrier:} block tree, epoch leaders, votes, notarizations, and ancestry;
\item \textbf{legitimacy:} quorum notarization for a block;
\item \textbf{horizon:} a short visible chain of notarized blocks;
\item \textbf{collapse:} finalize the block justified by the notarized-chain rule;
\item \textbf{discarded distinctions:} losing leader proposals and non-notarized forks;
\item \textbf{repair mechanism:} continue epochs until the notarized-chain horizon satisfies finalization.
\end{itemize}

Streamlet expresses the same notarized-chain pattern with a short chain rule: votes produce notarizations, and a brief notarized chain is enough to finalize a prefix.

\subsection{Algorand}

Algorand uses cryptographic sortition to choose small committees for Byzantine agreement on each block \cite{gilad-algorand-2017}. The communication carrier therefore includes not only proposals and votes, but also verifiable random function evidence showing that a participant was eligible to speak in a particular step. Collapse is mediated by committee certificates rather than a fixed all-validator quorum.

\paragraph{Protocol vocabulary.}
VRF sortition, committee, credential, proposer, BA step, soft vote, certify vote, block certificate.

\paragraph{Collapse-policy specification.}
\begin{specdisplay}
\mathcal{H}_{\mathrm{Algorand}} &:& \specdesc{block proposals, VRF credentials, committee votes,\\ steps, certificates} \\
\mathcal{E}_{\mathrm{Algorand}} &:& \specdesc{eligible committee votes\\ with valid sortition evidence} \\
\mathcal{V}_{\mathrm{Algorand}} &:& \specdesc{round and BA-step committee horizon} \\
\mathcal{P}_{\mathrm{Algorand}} &:& \text{certified block for a round} \\[0.4em]
L_{\mathrm{Algorand}}(h) &\mathrel{:=}&
\{(v,b,s,c) \mid c \text{ proves voter } v \text{ is eligible in step } s \text{ and votes for } b\} \\
R_{\mathrm{Algorand},r,s}(e) &\mathrel{:=}&
\{(v,b,s',c) \in e \mid s'=s \text{ and the vote belongs to round } r\} \\
C_{\mathrm{Algorand}}(u) &\mathrel{:=}&
\begin{cases}
b & \text{if } u \text{ contains sufficient eligible committee votes certifying } b,\\
\text{undefined} & \text{otherwise.}
\end{cases}
\end{specdisplay}

\paragraph{Protocol reading.}
VRF credentials make committee membership portable evidence. The round and step define the horizon, committee votes supply decision-grade evidence, and certification collapses the visible committee history into a block decision.

\paragraph{Comparative reading.}
In the normal form:
\begin{itemize}[leftmargin=1.2em]
\item \textbf{carrier:} proposed blocks, VRF credentials, committee votes, BA steps, and certificates;
\item \textbf{legitimacy:} valid sortition evidence and sufficient eligible committee support;
\item \textbf{horizon:} one round and its BA-step committee views;
\item \textbf{collapse:} certify one block for the round and extend the prefix;
\item \textbf{discarded distinctions:} non-selected users, losing proposals, and committee votes outside the certified block path;
\item \textbf{repair mechanism:} advance BA steps or rounds when the current committee horizon does not certify a block.
\end{itemize}

Algorand shifts the carrier into privately selected committees. Instead of routing evidence through a stable leader or broad validator set, the protocol carries eligibility proof with each vote and collapses each BA step through committee certificates.

\subsection{Ouroboros}

Ouroboros is the proof-of-stake longest-chain branch of the blockchain consensus lineage \cite{kiayias-ouroboros-2017,david-ouroboros-praos-2018}. It does not expose finality through the same immediate quorum-certificate interface as HotStuff or Tendermint. Instead, slot leaders extend a block tree, validators apply a chain-selection rule, and the stable prefix of the selected chain becomes the terminal artifact.

\paragraph{Protocol vocabulary.}
Slot, epoch, stake distribution, slot leader, leader eligibility, block tree, chain-selection rule, stability parameter, common prefix.

\paragraph{Collapse-policy specification.}
\begin{specdisplay}
\mathcal{H}_{\mathrm{Ouroboros}} &:& \specdesc{slot-indexed block tree, leader evidence, stake distribution,\\ local chain views} \\
\mathcal{E}_{\mathrm{Ouroboros}} &:& \specdesc{valid blocks with eligible-leader\\ and ancestry evidence} \\
\mathcal{V}_{\mathrm{Ouroboros}} &:& \specdesc{locally selected best chain with a stability window} \\
\mathcal{P}_{\mathrm{Ouroboros}} &:& \text{stable chain prefix} \\[0.4em]
L_{\mathrm{Ouroboros}}(h) &\mathrel{:=}&
\{b \in \mathrm{blocks}(h) \mid b \text{ has valid slot, leader, stake, and parent evidence in } h\} \\
R_{\mathrm{Ouroboros},k}(e) &\mathrel{:=}&
\mathrm{bestChain}(e) \text{ together with the } k\text{-slot or } k\text{-block stability horizon} \\
C_{\mathrm{Ouroboros},k}(v) &\mathrel{:=}&
\mathrm{prefix}_k(v),
\end{specdisplay}
where \(\mathrm{prefix}_k(v)\) means the selected chain with its unstable suffix removed.

\paragraph{Protocol reading.}
The block tree is the carrier, leader eligibility and block validity supply legitimacy, the chain-selection rule chooses the visible chain, and the stability parameter determines which prefix is old enough to expose as final for the protocol's security claim.

\paragraph{Comparative reading.}
In the normal form:
\begin{itemize}[leftmargin=1.2em]
\item \textbf{carrier:} slot-indexed block tree, stake distribution, leader evidence, block headers, and local chain views;
\item \textbf{legitimacy:} valid leader election or eligibility evidence, valid ancestry, and stake assumptions;
\item \textbf{horizon:} chain-selection view plus a stability window;
\item \textbf{collapse:} expose the stable prefix of the selected chain;
\item \textbf{discarded distinctions:} losing forks, withheld blocks, and unstable suffix alternatives outside the stable prefix;
\item \textbf{repair mechanism:} continue extending the block tree and recompute the selected chain until the common-prefix horizon makes earlier blocks stable.
\end{itemize}

Ouroboros makes stable-prefix projection explicit over a stake-selected chain/tree. It does not require every block to carry a direct finality certificate; instead, the selected chain is exposed through a stability window under the protocol's stake and synchrony assumptions.

\subsection{GRANDPA}

GRANDPA is a finality gadget: it finalizes prefixes of a block tree produced by an underlying block-production mechanism \cite{stewart-grandpa-2020}. Its communication carrier is not just the produced chain. It includes validator prevotes, precommits, rounds, and ancestry relations, so votes for descendants also count as evidence for ancestors.

\paragraph{Protocol vocabulary.}
Finality gadget, round, prevote, precommit, estimate, block tree, ancestry, supermajority, justification, finalized prefix.

\paragraph{Collapse-policy specification.}
\begin{specdisplay}
\mathcal{H}_{\mathrm{GRANDPA}} &:& \specdesc{block tree, rounds, prevotes, precommits, voter weights,\\ ancestry} \\
\mathcal{E}_{\mathrm{GRANDPA}} &:& \specdesc{weighted vote evidence over blocks\\ and descendants} \\
\mathcal{V}_{\mathrm{GRANDPA}} &:& \specdesc{round-local vote graph projected through ancestry} \\
\mathcal{P}_{\mathrm{GRANDPA}} &:& \text{finalized block prefix} \\[0.4em]
L_{\mathrm{GRANDPA}}(h) &\mathrel{:=}&
\{(v,b,\phi,r) \mid v \text{ casts valid } \phi \in \{\mathrm{prevote},\mathrm{precommit}\} \text{ for } b \text{ in round } r\} \\
R_{\mathrm{GRANDPA},r}(e) &\mathrel{:=}&
\{(v,b,\phi,r') \in e \mid r'=r \text{ and } b \text{ is interpreted through block ancestry}\} \\
C_{\mathrm{GRANDPA}}(u) &\mathrel{:=}&
\begin{cases}
b & \begin{array}[t]{@{}l@{}}
\text{if } b \text{ is the highest block justified by}\\
\text{supermajority precommit evidence in } u,
\end{array}\\
\text{undefined} & \text{otherwise.}
\end{cases}
\end{specdisplay}

\paragraph{Protocol reading.}
GRANDPA separates block production from finality. The block tree supplies the candidate carrier, validator votes supply decision-grade evidence, ancestry lifts votes on descendants to support ancestors, and a round finalizes the highest block whose prefix is justified by sufficient precommit evidence.

\paragraph{Comparative reading.}
In the normal form:
\begin{itemize}[leftmargin=1.2em]
\item \textbf{carrier:} block tree, validator votes, rounds, weights, ancestry, and finality justifications;
\item \textbf{legitimacy:} weighted supermajority prevote/precommit evidence;
\item \textbf{horizon:} a GRANDPA round and the ancestry closure of votes in that round;
\item \textbf{collapse:} finalize the highest block whose prefix is justified by supermajority precommit evidence;
\item \textbf{discarded distinctions:} competing block-tree branches outside the finalized prefix;
\item \textbf{repair mechanism:} move to later rounds with updated estimates when the current round cannot justify a new prefix.
\end{itemize}

GRANDPA leaves block production as the source of candidate history and uses weighted prevote/precommit evidence to decide when a prefix of that history can stop being provisional.

\subsection{Casper FFG}

Casper the Friendly Finality Gadget is another overlay finality mechanism \cite{buterin-casper-ffg-2017}. It organizes finality around checkpoint votes. Validators vote for source-target checkpoint links, and justified checkpoints can become finalized when later votes extend them in the required way. The carrier is therefore a checkpoint tree plus validator votes and slashing-relevant evidence.

\paragraph{Protocol vocabulary.}
Checkpoint, epoch, source, target, vote, justification, finalization, slashing condition, validator deposit.

\paragraph{Collapse-policy specification.}
\begin{specdisplay}
\mathcal{H}_{\mathrm{FFG}} &:& \specdesc{checkpoint tree, source-target votes, validator weights,\\ slashing evidence} \\
\mathcal{E}_{\mathrm{FFG}} &:& \specdesc{weighted checkpoint-link evidence and\\ justified-checkpoint evidence} \\
\mathcal{V}_{\mathrm{FFG}} &:& \specdesc{epoch-local checkpoint horizon} \\
\mathcal{P}_{\mathrm{FFG}} &:& \text{finalized checkpoint prefix} \\[0.4em]
L_{\mathrm{FFG}}(h) &\mathrel{:=}&
\{(v,s,t) \mid v \text{ casts a valid vote from source checkpoint } s \text{ to target } t\} \\
R_{\mathrm{FFG},e}(x) &\mathrel{:=}&
\{(v,s,t) \in x \mid t \text{ is visible in epoch horizon } e\} \\
C_{\mathrm{FFG}}(u) &\mathrel{:=}&
\begin{cases}
t & \text{if } u \text{ justifies and finalizes checkpoint } t,\\
\text{undefined} & \text{otherwise.}
\end{cases}
\end{specdisplay}

\paragraph{Protocol reading.}
Checkpoint votes are the communication evidence, stake weight supplies legitimacy, and slashing conditions make conflicting evidence costly. Collapse happens when the checkpoint graph contains enough weighted votes to justify and then finalize a checkpoint prefix.

\paragraph{Comparative reading.}
In the normal form:
\begin{itemize}[leftmargin=1.2em]
\item \textbf{carrier:} checkpoint tree, validator votes, source-target links, stake weights, and slashing evidence;
\item \textbf{legitimacy:} weighted supermajority votes respecting slashing conditions;
\item \textbf{horizon:} an epoch-local view of justified checkpoints and target votes;
\item \textbf{collapse:} finalize a checkpoint and expose the prefix below it;
\item \textbf{discarded distinctions:} competing checkpoint branches and votes outside the finalized prefix;
\item \textbf{repair mechanism:} continue voting in later epochs until checkpoint evidence justifies and finalizes a prefix.
\end{itemize}

Casper FFG separates block production from finality through checkpoint links and slashing-aware validator votes. Its collapse horizon is a checkpoint tree rather than GRANDPA's round-local prevote/precommit ancestry rule.

\subsection{Snowman and Avalanche}

Avalanche-style consensus uses repeated random sampling to drive local preferences into a metastable common choice \cite{rocket-avalanche-2019}. Snowman specializes the family to a linear-chain setting. The communication carrier is a sequence of sampled preference observations and confidence counters for competing chain choices, and the chain choice becomes stable once the confidence threshold is met.

\paragraph{Protocol vocabulary.}
Repeated sampling, preference, query, response, confidence counter, metastability, block conflict, Snowman chain.

\paragraph{Collapse-policy specification.}
\begin{specdisplay}
\mathcal{H}_{\mathrm{Snowman}} &:& \specdesc{block proposals, sampled responses, preferences, conflicts,\\ ancestry, confidence counters} \\
\mathcal{E}_{\mathrm{Snowman}} &:& \specdesc{sample-majority evidence for a block\\ or chain extension} \\
\mathcal{V}_{\mathrm{Snowman}} &:& \specdesc{local sampling window for a conflicting block choice} \\
\mathcal{P}_{\mathrm{Snowman}} &:& \text{accepted chain prefix} \\[0.4em]
L_{\mathrm{Snowman}}(h) &\mathrel{:=}&
\{(b,m) \mid m \text{ is sampled support for block } b \text{ or its chain ancestry in } h\} \\
R_{\mathrm{Snowman},\beta}(e) &\mathrel{:=}&
\{(b,m) \in e \mid b \text{ has visible sample support contributing to confidence threshold } \beta\} \\
C_{\mathrm{Snowman}}(u) &\mathrel{:=}&
\begin{cases}
b & \text{if } b \text{ reaches the required confidence threshold in } u,\\
\text{undefined} & \text{otherwise.}
\end{cases}
\end{specdisplay}

\paragraph{Protocol reading.}
Sampling responses are the evidence carrier. The horizon is not a view, round certificate, or finality checkpoint, but a confidence threshold over repeated local observations for a conflicting block choice. Collapse occurs when one block preference has accumulated enough sampled support to extend the accepted chain prefix.

\paragraph{Comparative reading.}
In the normal form:
\begin{itemize}[leftmargin=1.2em]
\item \textbf{carrier:} sampled queries and responses, block preferences, conflict sets, confidence counters, and chain ancestry;
\item \textbf{legitimacy:} sample-majority support under Snowman's sampling and adversary assumptions;
\item \textbf{horizon:} a local confidence window and threshold for a block choice;
\item \textbf{collapse:} accept a chain extension once confidence is high enough;
\item \textbf{discarded distinctions:} losing block preferences and sampled histories that no longer affect the accepted prefix;
\item \textbf{repair mechanism:} continue sampling until confidence accumulates or preferences shift.
\end{itemize}

Snowman and the broader Avalanche family use repeated randomized sampling instead of quorum intersection, certificate chains, or finality votes. The carrier grows as sampled observations update local preference and confidence order; collapse happens when that order has a stable maximal choice.

\subsection{Jolteon and Ditto}

Jolteon and Ditto sharpen the HotStuff lineage by making network adaptation explicit. Jolteon optimizes the fast path under favorable network behavior, while Ditto adds an asynchronous fallback path when the optimistic path cannot safely make progress \cite{gelashvili-jolteon-ditto-2022}.

\paragraph{Protocol vocabulary.}
Fast path, fallback path, proposal, vote, quorum certificate, timeout certificate, view, certified chain.

\paragraph{Collapse-policy specification.}
\begin{specdisplay}
\mathcal{H}_{\mathrm{JD}} &:& \specdesc{proposals, votes, QCs, timeout/fallback evidence,\\ block ancestry} \\
\mathcal{E}_{\mathrm{JD}} &:& \specdesc{certified proposal-chain and fallback evidence} \\
\mathcal{V}_{\mathrm{JD}} &:& \specdesc{optimistic view horizon or asynchronous fallback horizon} \\
\mathcal{P}_{\mathrm{JD}} &:& \text{committed block prefix} \\[0.4em]
L_{\mathrm{JD}}(h) &\mathrel{:=}&
\{(b,q) \mid q \text{ is a QC, timeout certificate, or fallback certificate for } b \text{ in } h\} \\
R_{\mathrm{JD},v}(e) &\mathrel{:=}&
\{(b,q) \in e \mid (b,q) \text{ is visible to the fast path or fallback path in view } v\} \\
C_{\mathrm{JD}}(u) &\mathrel{:=}&
\begin{cases}
\mathrm{commitFast}(u) & \text{if } u \text{ satisfies the optimistic commit rule},\\
\mathrm{commitFallback}(u) & \text{if } u \text{ satisfies the fallback commit rule},\\
\text{undefined} & \text{otherwise.}
\end{cases}
\end{specdisplay}

\paragraph{Protocol reading.}
The fast path uses a narrow leader/view horizon when timely evidence is available. The fallback path widens or changes the horizon when the optimistic collapse rule cannot safely produce a prefix.

\paragraph{Comparative reading.}
In the normal form:
\begin{itemize}[leftmargin=1.2em]
\item \textbf{carrier:} proposals, votes, quorum certificates, timeout/fallback evidence, and block ancestry;
\item \textbf{legitimacy:} quorum certificates and fallback certificates that carry decision-grade evidence across views;
\item \textbf{horizon:} the optimistic leader/view path when timely, or the fallback horizon when synchrony assumptions are not currently useful;
\item \textbf{collapse:} the fast path commits through certified chains when the network cooperates;
\item \textbf{discarded distinctions:} failed fast-path attempts and uncertified branches once a fallback or later certified path commits;
\item \textbf{repair mechanism:} asynchronous fallback preserves progress when the short optimistic horizon is insufficient.
\end{itemize}

Jolteon/Ditto makes horizon selection adaptive. The protocol does not choose once between ``synchronous'' and ``asynchronous'' consensus. It changes the collapse policy depending on whether the fast horizon currently carries enough evidence.

\subsection{DAG-Rider}

DAG-Rider first builds a reliable-broadcast DAG and then interprets it in waves. The carrier is the local DAG of reliably broadcast vertices, round edges, and parent references. The legitimacy relation is reliable-broadcast validity plus sufficient round support, typically a threshold-sized set of vertices. The horizon is the wave: a bounded set of rounds used for local interpretation.

The collapse step selects or interprets leaders in the wave and emits an ordered sequence. If the wave does not yield progress, the protocol does not discard the DAG carrier. Later waves and randomized leader selection are the repair path, providing expected progress while preserving already disseminated causal evidence. The DAG is not transport noise; it is the object the ordering rule reads.

\paragraph{Protocol vocabulary.}
Reliable broadcast, DAG, vertex, round, wave, leader selection.

\paragraph{Collapse-policy specification.}
\begin{specdisplay}
\mathcal{H}_{\mathrm{DAG\text{-}Rider}} &:& \specdesc{reliable-broadcast DAG, vertices, rounds,\\ parent references} \\
\mathcal{E}_{\mathrm{DAG\text{-}Rider}} &:& \specdesc{RB-valid vertices with threshold round support} \\
\mathcal{V}_{\mathrm{DAG\text{-}Rider}} &:& \specdesc{wave-local DAG evidence} \\
\mathcal{P}_{\mathrm{DAG\text{-}Rider}} &:& \text{ordered output sequence} \\[0.4em]
L_{\mathrm{DAG\text{-}Rider}}(h) &\mathrel{:=}&
\{x \in \mathrm{vertices}(h) \mid x \text{ is RB-valid and threshold-supported}\} \\
R_{\mathrm{DAG\text{-}Rider},w}(e) &\mathrel{:=}&
\{x \in e \mid x \text{ lies in wave } w\} \\
C_{\mathrm{DAG\text{-}Rider}}(u) &\mathrel{:=}&
\mathrm{order}(\mathrm{leaders}(u),u).
\end{specdisplay}

\paragraph{Protocol reading.}
The DAG is the carrier, reliable-broadcast validity and round support are the legitimacy witnesses, the wave is the horizon, and later waves plus randomness serve as the repair path for progress.

\paragraph{Comparative reading.}
DAG-Rider separates dissemination from ordering by first building a reliable-broadcast DAG and then locally interpreting the DAG in waves \cite{keidar-dagrider-2021}. In the normal form:
\begin{itemize}[leftmargin=1.2em]
\item \textbf{carrier:} a round-based DAG of reliably broadcast vertices;
\item \textbf{legitimacy:} sufficient round evidence, typically $2f+1$ vertices, and reliable broadcast validity;
\item \textbf{horizon:} waves over the DAG;
\item \textbf{collapse:} local wave interpretation and random leader selection produce a total order output;
\item \textbf{discarded distinctions:} DAG concurrency and unchosen wave alternatives after the output order is emitted;
\item \textbf{repair mechanism:} randomized leader selection and later waves provide expected progress in asynchrony.
\end{itemize}

DAG-Rider separates the two layers cleanly: build communication history first, interpret it second. The reliable-broadcast DAG is not a transport detail; it is the object read by the wave rule.

\subsection{Hashgraph}

Hashgraph builds a signed event DAG by gossip-about-gossip. Each event records transactions plus hashes of a self-parent and an other-parent, so the graph records the causal path by which gossip spread. The interpretation layer assigns rounds, identifies witnesses, decides witness fame by virtual voting, and then orders events by receipt round and consensus timestamp.

The horizon is not a leader view or a fixed wave selected by an external coordinator. It is the witness-round evidence visible in the hashgraph. If a witness's fame is not yet decided, the protocol does not force an order from the current cut. Later gossip extends the carrier, and later rounds provide the virtual-vote evidence needed to decide fame.

\paragraph{Protocol vocabulary.}
Event, self-parent, other-parent, gossip-about-gossip, seeing, strongly seeing, witness, famous witness, virtual voting, consensus timestamp.

\paragraph{Collapse-policy specification.}
\begin{specdisplay}
\mathcal{H}_{\mathrm{Hashgraph}} &:& \specdesc{signed events with self-parent and other-parent edges} \\
\mathcal{E}_{\mathrm{Hashgraph}} &:& \specdesc{witness, seeing, strongly seeing, and fame evidence} \\
\mathcal{V}_{\mathrm{Hashgraph}} &:& \specdesc{witness-round horizon used for virtual voting} \\
\mathcal{P}_{\mathrm{Hashgraph}} &:& \text{total order of events or transactions} \\[0.4em]
L_{\mathrm{Hashgraph}}(h) &\mathrel{:=}&
\left\langle
\begin{array}{l}
\{w \in \mathrm{events}(h) \mid w \text{ is a witness}\},\\
\{(x,w) \mid x \text{ strongly sees witness } w \text{ in } h\}
\end{array}
\right\rangle \\
R_{\mathrm{Hashgraph},r}(e) &\mathrel{:=}&
\text{the witness evidence in and after round } r \text{ used to decide fame} \\
C_{\mathrm{Hashgraph}}(u) &\mathrel{:=}&
\mathrm{order}(\mathrm{famous}(u),u).
\end{specdisplay}

\paragraph{Protocol reading.}
The signed event DAG is the carrier, strongly seeing relations provide the supermajority evidence, witness rounds form the horizon, and virtual voting collapses the DAG into a total order once famous witnesses and receipt information are determined.

\paragraph{Comparative reading.}
Hashgraph also makes communication history explicit, but with a different interpretation layer \cite{baird-hashgraph-2016}. Participants gossip signed events, and each event names a self-parent and an other-parent. The resulting hashgraph records not only transactions, but who learned what from whom and when. Consensus is then computed by virtual voting over this shared causal structure rather than by sending separate vote messages.

In the normal form:
\begin{itemize}[leftmargin=1.2em]
\item \textbf{carrier:} signed hashgraph events with self-parent and other-parent edges, transactions, timestamps, and creator identities;
\item \textbf{legitimacy:} supermajority evidence expressed through seeing, strongly seeing, witnesses, and famous witnesses;
\item \textbf{horizon:} rounds of witnesses and later witness evidence used to decide fame;
\item \textbf{collapse:} virtual voting decides famous witnesses, then receipt rounds and consensus timestamps produce a total order;
\item \textbf{discarded distinctions:} event-DAG concurrency and non-famous witness alternatives after the consensus order is emitted;
\item \textbf{repair mechanism:} continued gossip extends the hashgraph until later rounds contain enough virtual-vote evidence to decide fame and order.
\end{itemize}

In Hashgraph, gossip-about-gossip builds an order-2 record of communication. Virtual voting reads that record as evidence and collapses it into an order-1 total order only after witness fame and receipt information are determined.

\subsection{Aleph}

Aleph is another explicit DAG/partial-order instance of the carrier/projection pattern \cite{gagol-aleph-2019}. It constructs a partially ordered set of units and then computes a common extension to a total order using local interpretation and randomized common-coin machinery. In this respect it connects Hashgraph's gossip-about-gossip style with the later DAG-BFT protocols: the communication carrier is a partial order, and the terminal artifact is a linear order of messages or transactions.

\paragraph{Protocol vocabulary.}
Unit, partial order, fork, creator, level, common coin, randomness beacon, total-order extension.

\paragraph{Collapse-policy specification.}
\begin{specdisplay}
\mathcal{H}_{\mathrm{Aleph}} &:& \specdesc{units/messages with creator identities and dependency edges} \\
\mathcal{E}_{\mathrm{Aleph}} &:& \specdesc{validated units, fork evidence, threshold support,\\ randomness evidence} \\
\mathcal{V}_{\mathrm{Aleph}} &:& \specdesc{locally visible poset layer plus coin evidence} \\
\mathcal{P}_{\mathrm{Aleph}} &:& \text{total order of units or transactions} \\[0.4em]
L_{\mathrm{Aleph}}(h) &\mathrel{:=}&
\{u \in \mathrm{units}(h) \mid u \text{ is valid and not excluded by fork evidence in } h\} \\
R_{\mathrm{Aleph},\ell}(e) &\mathrel{:=}&
\{u \in e \mid u \text{ is visible in layer/horizon } \ell \text{ with the relevant coin evidence}\} \\
C_{\mathrm{Aleph}}(v) &\mathrel{:=}&
\mathrm{extendOrder}(v)
\quad\text{when } v \text{ determines the next common total-order fragment.}
\end{specdisplay}

\paragraph{Protocol reading.}
The partial order of units is the carrier, validation and fork evidence determine admissible units, and common-coin evidence helps validators compute the same total-order extension from their local copies.

\paragraph{Comparative reading.}
In the normal form:
\begin{itemize}[leftmargin=1.2em]
\item \textbf{carrier:} units/messages arranged in a locally observed partial order, with creator identity and dependency edges;
\item \textbf{legitimacy:} validation of units, fork/equivocation checks, threshold support, and randomness-beacon evidence;
\item \textbf{horizon:} locally visible layers of the partial order together with common-coin decisions used for ordering;
\item \textbf{collapse:} a common deterministic interpretation, aided by randomness, extends the partial order into a total order;
\item \textbf{discarded distinctions:} concurrency among incomparable units after the final total order is emitted;
\item \textbf{repair mechanism:} continue extending the poset and invoking randomized ordering steps until enough common evidence is available.
\end{itemize}

Aleph first builds an order-2 poset carrier and only later computes an order-1 atomic-broadcast sequence. Randomness helps local interpretations converge on the same extension when the asynchronous carrier has enough evidence.

\subsection{Narwhal, Tusk, and Bullshark}

Narwhal and Tusk split data availability and causal dissemination into a mempool layer. Bullshark keeps the DAG shape but adds deterministic commit rules under partial synchrony. The carrier is the certified DAG block graph: batches, certificates, and parent references. The legitimacy relation is certification of blocks and sufficient parent support, which makes data availability and ancestry portable.

The horizon is a round or commit-rule window over certified DAG structure. Collapse maps the available DAG into ordered output, either through selected leaders, Tusk-style local interpretation, or Bullshark-style deterministic rules. If a block is available but not yet orderable, the protocol preserves it for later rules instead of forcing an early prefix decision. Dissemination and order are decoupled, but the DAG still carries the evidence that later ordering consumes.

\paragraph{Protocol vocabulary.}
Narwhal, Tusk, Bullshark, mempool, data availability, causal histories, parent support, commit rules.

\paragraph{Collapse-policy specification.}
\begin{specdisplay}
\mathcal{H}_{\mathrm{NTB}} &:& \specdesc{certified DAG blocks, batches, causal references,\\ mempool data} \\
\mathcal{E}_{\mathrm{NTB}} &:& \specdesc{block certificates and parent-support evidence} \\
\mathcal{V}_{\mathrm{NTB}} &:& \specdesc{round or commit-rule window} \\
\mathcal{P}_{\mathrm{NTB}} &:& \text{ordered output} \\[0.4em]
L_{\mathrm{NTB}}(h) &\mathrel{:=}&
\{b \in \mathrm{blocks}(h) \mid b \text{ is certified and parent-supported}\} \\
R_{\mathrm{NTB},w}(e) &\mathrel{:=}&
\{b \in e \mid b \text{ is visible to commit window } w\} \\
C_{\mathrm{NTB}}(u) &\mathrel{:=}&
\{b \in u \mid \mathrm{dagRule}(u,b) \text{ orders } b\}.
\end{specdisplay}

\paragraph{Protocol reading.}
Dissemination and causal history are carried in the certified DAG, certification and parent support are the legitimacy witnesses, and the commit-rule window determines when preserved evidence becomes orderable.

\paragraph{Comparative reading.}
Narwhal and Tusk separate the DAG mempool from consensus, making data availability and causal dissemination a first-class layer \cite{danezis-narwhal-tusk-2022}. Bullshark continues the DAG-BFT line with deterministic commit rules under partial synchrony \cite{spiegelman-bullshark-2022}. In the normal form:
\begin{itemize}[leftmargin=1.2em]
\item \textbf{carrier:} certified DAG blocks carrying batches and references to prior blocks;
\item \textbf{legitimacy:} certificates over DAG blocks and sufficient parent support;
\item \textbf{horizon:} rounds and commit rules over certified DAG structure;
\item \textbf{collapse:} selected leaders or commit rules map DAG availability into ordered output;
\item \textbf{discarded distinctions:} concurrent data availability relations once a final order is exposed;
\item \textbf{repair mechanism:} DAG accumulation lets later commit rules use evidence that was not ready at an earlier output point.
\end{itemize}

For Narwhal, Tusk, and Bullshark, dissemination history is not mere transport. It is the certified DAG object from which ordering decisions are derived.

\end{document}